\documentclass[12pt,leqno]{amsart}
 
\usepackage{amsmath}
\usepackage{graphicx}
\input amssym.def
\input amssym

\long\def\symbolfootnote[#1]#2{\begingroup%
\def\thefootnote{\fnsymbol{footnote}}\footnote[#1]{#2}\endgroup}

\newtheorem{theorem}{\sc Theorem}[section] 
\newtheorem{lemma}[theorem]{\noindent {\sc Lemma}} 
\newtheorem{corollary}[theorem]{\sc Corollary}
\newtheorem{proposition}[theorem]{\sc Proposition}

\newtheorem{definition}{Definition}[theorem]

\theoremstyle{plain}

\renewcommand{\a}{\alpha}
\renewcommand{\b}{\beta}
\renewcommand{\d}{\delta}

\newcommand{\e}{\varepsilon}

\newcommand{\g}{\gamma}
\newcommand{\G}{\Gamma}
\renewcommand{\l}{\lambda}

\renewcommand{\L}{\Lambda}

\newcommand{\s}{\sigma}

\renewcommand{\t}{\tau}
\newcommand{\cal}{\mathcal}

\newcommand{\R}{{\Bbb R}}
\newcommand{\C}{{\Bbb C}}
\newcommand{\F}{{\Bbb F}}
\renewcommand{\O}{\Omega}
\renewcommand{\o}{\omega}

\renewcommand{\i}{\infty}
\newcommand{\p}{\partial}

\renewcommand{\thefootnote}{\fnsymbol{footnote}}
\newcommand{\nat}{\natural}
\renewcommand{\thefootnote}{\fnsymbol{footnote}}
\addtocontents{toc}{\hfill Page\endgraf}
\addtocontents{lof}{\ Figure\hfill Page}

\makeatletter
\renewcommand\tableofcontents{%
  \newfont{\scaledfont}{cmr12 scaled 7000}
  \parindent\z@\raggedright
  \scaledfont\contentsname\par\normalsize%
  \rule{\textwidth}{1pt}
  \nobreak
  \vskip 40\p@
  \@starttoc{toc}%
}
\makeatother
\begin{document}
 \author[J. Harrison  Department of Mathematics  U.C. Berkeley]{J. Harrison
\\Department of Mathematics
\\University of California, Berkeley}
\title[Chainlets]{Differential complexes and exterior calculus}
\date{December 21,2005, edited May 12, 2006}

\begin{abstract} 
  In this paper we present a new theory of calculus over $k$-dimensional domains in a smooth $n$-manifold, unifying the discrete, exterior,  and continuum theories.     The calculus begins at a single point and is extended to chains of finitely many points by linearity, or superposition. It    converges to the smooth continuum with respect to a norm on the space of ``monopolar chains,'' culminating in the chainlet complex.  Through this complex, we discover a broad theory of coordinate free, multivector analysis in smooth manifolds for which both the classical Newtonian calculus and the Cartan exterior calculus become special cases. The chainlet operators, products and integrals   apply to both symmetric and antisymmetric tensor cochains.   As   corollaries, we obtain the full calculus on  Euclidean space, cell complexes, bilayer structures (e.g., soap films) and nonsmooth domains, with equal ease.   The power comes from the recently   discovered prederivative and preintegral that are antecedent to the Newtonian theory. These lead to new models for the continuum of space and time, and permit analysis of domains that may not be locally Euclidean, or locally connected, or with locally finite mass.   
  \end{abstract}
\maketitle
  
\section*{Preface}
 We put forward a novel meaning of the real continuum which is found by first developing a full theory of calculus at a single point -- the origin, say,  of a vector space -- then carrying  it over to domains supported in finitely many points in an affine space, and finally extending it to the class of ``chainlets''    found by taking limits of the discrete theory with respect to a norm.   Local Euclidean structure is not necessary for the calculus to hold.    The calculus extends to $k$-dimensional domains in $n$-manifolds.  We do not rely on any results or definitions of classical calculus to develop our theory.   In the appendix we show how to derive the standard results of single and multivariable calculus in Euclidean space as direct corollaries.

 This preprint is in draft form, sometimes rough.    There are some details which are still linked to earlier versions of the theory.   It is being expanded into a text  which includes new  applications, numerous examples, figures, exercises, and necessary background beyond basic linear algebra, none of which are included below.

\subsection*{Problems of the classical approach}
As much as we all love the calculus, there have been limits to our applications in both pure and applied mathematics coming from the definitions which arose during the ``rigorization'' period of calculus. Leibniz had searched for an 'algebra of geometry', but this was not available until Grassmann realized the importance of $k$-vectors  in his seminal 1844 paper.  But his ideas were largely ignored until Gibbs and Clifford began to appreciate them in 1877.  Cartan used the Grassmann algebra to develop the exterior calculus.  Meanwhile, Hamilton and others were debating the rigorization of multivariable calculus. It was not even clear what was meant by ``space'' in the late 19th century.  The notions of div, grad and curl of Gibbs and Heaviside, were finally settled upon by the bulk of the community, and there have been two separate, and often competing, approaches ever since.  But the Cartan theory, rather than the coordinate theory we teach our freshmen, has given mathematicians and physicists a clearer vision to guide great leaps of thought, and this theory has led to much of the mathematics behind the prizewinning discoveries of mathematics and mathematical physics.   The group of those who now understand the importance of the Cartan theory is growing, as evidenced by the many books and papers which now start with this theory as their basis. 

The reader might well ask, what are some of the problems of coordinate calculus?  Does this Cartan theory have limitations?  Why do we need something new?

First of all, the coordinate theory requires multiple levels of limits upon limits to be able to  understand interesting applications in manifolds or curved space that take years of training to understand.   We become proud virtuosos of our coordinate techniques.  But these methods are a barrier to others who do not have the time or patience to learn them.  And they also form a barrier to those of us who rely on them for they may cloud our vision with their complexity.  After we become experts in Euclidean space with our div, grad and curl operators, we then move into curved space.   Everything works quite well in restricted settings of smooth Riemannian manifolds and submanifolds. We can manipulate the limits,  know when to change their orders, and do term by term integration with Fourier series and wavelets, etc.   But when we try to reduce our assumptions and work with less smooth domains, permitting piecewise linear, corners, Lipschitz conditions, or introduce singularities, the coordinate theory becomes more and more intractable, and soon begins to break down. We are forced to consider too much information that is not necessary, and need more and more assumptions, and limits, to get anywhere. The classical calculus cannot treat everywhere nonsmooth domains as there are be no tangent spaces to work with.  The Cartan theory cannot handle important bilayer structures such as soap films because the boundary operator introduces extraneous terms along branched curves.   The standard Cartan proof to Stokes' theorem relies on boundaries matching and cancelling with opposite orientation where they meet so local connectivity and locally finite mass are required.  Discrete theories, so important today, also have problems.  It can be surprisingly difficult to be certain when supposed approximations actually do converge to something meaningful. Bootstrapping techniques still prevail.   Cochains in discrete or combinatorial theories usually fail to satisfy a basic property such as  commutativity or associativity of wedge product, or existence of a satisfactory Hodge star operator.    Simplicial complexes have appealing simplicity, but it is surprisingly difficult to model vector fields with them. There is a great deal of  information in a simplicial complex that is not needed for calculus.  There are corners, matching boundaries of simplices, ratios of length to area, and so forth, that must be considered.

As we try to apply calculus to important microscopic relations of physics or biology, the classical and discrete theories fail us.  Both the Cartan model and the coordinate model assume a rigid,
        locally Euclidean continuum.  This forces the calculus to completely break
       down in quantum mechanics.  There is simple mathematics going on that cannot be seen with an assumption of a local Euclidean structure with it its ``infinitesimal connectivity''.

\subsection*{Geometrization of Dirac delta functions} In this paper we present a new approach to calculus in which more efficient choices of limits are taken at key points of the development, greatly reducing the number of limits needed for the full theory. 
 This work is motivated by Grassmann's algebra of $k$-vectors $$\L = \oplus_k \L_k$$ in a vector space $V$, but takes his ideas further.  Instead of using $k$-vectors   which have ambiguous geometric meaning, we do something slightly different to recover geometric meaning.  We shrink a $k$-cell to the origin, say, renormalizing its mass at each stage so it remains constant.  The question is "what do you get in the limit?"  Intuitively, these geometrical ``infinitesimals'' are like ``mass points'', or geometric versions of Dirac delta points.   In this paper, we show that the limit exists in a normed space $\cal{N}_k^{\i}$ and depends only on the $k$-direction and mass of the $k$-cell.  We call the limit a {\itshape  $(0,k)$-pole} at the origin.  The space   of $(0,k)$-poles at the origin is  isomorphic to the space of exterior $k$-vectors $\L_k.$  The boundary operator   $\p: \cal{N}_k^{\i} \to \cal{N}_{k-1}^{\i}$  is continuous, and thus determines the boundary of a $(0,k)$-pole as a sum of    
 $(1,k-1)$-poles, a geometrization of Dirac dipoles of dimension $k-1$.   A geometric directional derivative operator $\nabla_u: \cal{N}_k^{\i} \to \cal{N}_{k}^{\i}$ is defined on $(1,k)$-poles, leading to geometrizations of quadrupoles of abritrary order.  We call $\nabla_u$ a {\itshape  prederivative}.  Its dual is the directional derivative of forms. The formal treatment begins with the use of the Koszul complex $X(V)$  in \S\ref{koszul}.  Our geometrical view is supported and  encapsulated by a matching algebraic construction of the Koszul complex which keeps track of all the implicit tensor algebra. For example, the Koszul complex shows that the boundary of a $(0,k)$-pole is well defined, and leads to a nice proof that $\p^2 = 0$.

 \subsection*{The real continuum beyond numbers}
  Arising from this theory are new models for the real continuum. Morris Hirsch wrote,\symbolfootnote[2]{An e-mail message, quoted with permission}     
  on December 13, 2003.  \begin{quotation} A basic philosophical problem has been to make sense of ``continuum'',
as
in the space
of real numbers, without introducing numbers.  Weyl   wrote,   ``The
introduction of numbers as coordinates ... is an act of violence''.
Poincar\'e wrote about the ``physical continuum'' of our intuition, as
opposed
to the mathematical continuum.  Whitehead (the philosopher) based our
use of
real numbers on our intuition of time intervals and spatial regions.
The
Greeks tried, but didn't get very far in doing geometry without real
numbers.
But no one, least of all the Intuitionists, has come up with even a
slightly
satisfactory replacement for basing the continuum on the real number
system,
or basing the real numbers on Dedekind cuts, completion of the
rationals, or
some equivalent construction. \end{quotation}
We propose chainlets as more flexible models for the real continuum, since our ``infinitesimal'' $(0,1)$-poles can be limits of intervals,  or equally well   intervals with countably many smaller intervals removed such as Cantor sets. 
The topology of  limiting approximations of $(0,1)$-poles does not matter, contrasted with $1$-dimensional tangent spaces which must have the structure of Euclidean space.   Moreover, we add layers of possibilities to the continuum model, by allowing higher order  poles at each point, a sum of a geometric monopole, dipole, quadrupole, etc., a kind of ``jet'' of geometry.  Philosophers have said ``there is only now'' in human experience, but this continuum model gives a model of time which includes in each moment a jet of time, something we might sense as a Gestalt experience when time seems nearly to come to a halt, or the opposite when time seems to rush by.  

   Our methods are different from those of nonstandard analysis which also defines ``infinitesimals,'' but where there is great effort to mimic the structure of the reals as closely as possible.  For the theory of hyperreals of nonstandard analysis to work, there is an ordering required, as well as inverses, transfer principles, etc. We only use local Euclidean space as an affine space to support our chainlets. We make use of the affine space structures of subtraction $x-y$ and existence of norms $|x-y|$.

\subsection*{Unification of viewpoints}
Mathematicians take many viewpoints in the pure theory and its applications, be they smooth manifolds, Lipschitz structures, polyhedra, fractals, finite elements, soap films, measures, numerical  methods, mathematical physics, etc.  The choice sets the stage and determines our audience and our methods.        All of these viewpoints can potentially be unified in chainlet theory.

Three basic theorems for chainlet domains $J$ lead to much of the classical theory, forming a  ``tripod'' of calculus.   

\begin{enumerate}  
\item $\int_{\p J} w = \int_J d w$  (Stokes' theorem)
\item $\int_{f_*J} w = \int_J f^* w$ (Change of variables)
\item $\int_{\perp J} w = \int_J * w$   (Star theorem)
\end{enumerate}

Each of these is optimal, each has a one line proof, after the initial definitions and basic continuity results.     (See \S\ref{chainletcomplex} and \cite{earlyhodge}.)    Only the Star theorem requires a metric. With a metric, (i) and (ii) imply a general divergence theorem
$$\int_{\perp \p J} \o = \int_J d \star \o,$$ and a general curl theorem  
$$\int_{\p \perp J}\o = \int_J \star d \o.$$   In the standard approach, one proves the Fundamental Theorem of Calculus first, and then eventually builds up to a general Stokes' theorem for manifolds.    We start with the three results above, as first results.

 In \S\ref{applications} we provide a general method for testing whether a dual triangulation will converge to the Hodge star smooth continuum as the mesh size tends to zero. For example, this method, along with our change of variables result \ref{change} and \ref{oldintegral}, can be used to show that the circumcentric dual introduced by  by Marsden, et al, \cite{marsden} does converge to the smooth continuum. (See \S\ref{applications} for a general method.)   The work in this paper, especially the continuity theorems for operators, products and relations in \S\ref{pointed}, should fill in the missing ingredients needed to show which of the operators, products and relations in the above works converge to the smooth continuum.

Besides the discrete theory, the author has developed two other extensions of calculus: 
\begin{enumerate}  
         \item {\bf   Bilayer calculus} with applications to the calculus of variations including Plateau's problem (soap bubbles) (See \cite{soap}, \cite{plateau}.)
  \item { \bf Calculus on nonsmooth domains} (e.g., fractals) (See \cite{HN1}, \cite{HN2}, \cite{stokes}, \cite{continuity}, \cite{iso}, \cite{earlyhodge}.)
  \end{enumerate}
  \subsection*{Differential complexes}\label{introduction}
    
Let $\O_k^r$ denote the space of differential $k$-forms  of class $C^r$ on a manifold $M^n$.   Since the exterior derivative $d$ satisfies $d^2 = 0$ it follows that
$$\O_{k+r}^{0} \buildrel d \over \leftarrow  \cdots \buildrel d \over \leftarrow \O_{k +1}^{r-1}  \buildrel d \over \leftarrow  \O_k^r \buildrel d \over \leftarrow \O_{k-1}^{r+1} \buildrel d \over \leftarrow \cdots \buildrel d \over \leftarrow \O_0^{k+r}$$
  is a chain complex for each $k+r = c \ge 1$. Versions of this complex provide the basis to much of analysis and topology.  Differential forms provide coordinate free integrands, leading to the Cartan exterior calculus \cite{cartan}.  De Rham cohomology theory is based on this complex.  Classical operators such as Hodge star and pullback are closed in this complex, leading to broad applications.    
   
Mathematicians have sought a matching differential covariant complex for domains, with $\p$ playing the role of $d$ and for which the complex is closed under smooth mappings.    Poincar\'e introduced the simplicial complex with its boundary operator mapping a simplicial complex of dimension $k$ into one of dimension $k-1.$      Many papers continue to be written from the viewpoint of the simplicial complex, especially in numerical analysis.  Although much progress has been made, (e.g. \cite{marsden}), the simplicial complex brings with it inherent problems.  The smooth pushforward operator is not closed in the simplicial complex, making iterative methods problematic.   Difficulties include the commutative cochain problem, lack of associativity of cochains, lack of natural definitions of vector fields, a well-behaved discrete Hodge star operator, and  convergence of operators and relations to the smooth continuum.    We know of no other discrete theory, apart from that presented in this paper, which solves all of these problems simultaneously.  For example, wedge product is defined for our discrete cochains as restriction of the wedge product for forms at discrete points and is commutative, by definition.  
 
 Beyond the needs of the discrete community, though, pure mathematicians  have sought a geometrically based, covariant differential complex for which  operators of analysis      act continuously on the complex. Of especial interest are operators  dual to operators on differential forms such as Hodge star $\star$, and those which commute with the pushforward operator, for these lead to well defined integral relations in manifolds.  An important goal has been to extend the Gauss divergence theorem to nonsmooth domains. (See Theorem \ref{div}.)
  
 Whitney \cite{whitney} introduced the vector space of polyhedral chains which does form a chain complex satisfying $\p^2 = 0$.    The Banach space of his sharp norm    is not a chain complex because the boundary operator on polyhedral chains is not continuous.    His flat norm does yield a chain complex with a well defined boundary operator.   This has led to applications in geometric measure theory \cite{federer}.  However, the  Hodge star operator is not closed in the flat covariant complex.   The flat norm  has no divergence theorem, as can be seen by the example in Whitney of a flat form $\o$ in $\R^2$ without flat components (\cite{whitney}, p. 270).  
 $$\o = \begin{cases}dx +dy, &x > -y \\0, & x \le -y\end{cases}.$$
 The net divergence of $\o$ is zero, yet the net flux across the boundary of a square with diagonal $x = -y$  is nonzero.  The problem resides with the Hodge star operator which is not continuous in the flat norm\footnote{More than one research group has not noticed this example, and tried to use the flat norm to develop calculus over fractal boundaries.}.   
 
 Schwarz's distributions and de Rham's currents   \cite{derham} solved the covariant chain complex problem almost too perfectly.   Currents ${\cal T}_k^r$ are defined to be continuous linear functionals on differential $k$-forms of class $C^r$ with compact support.
  Bounded operators on currents are defined by duality with bounded operators on forms.  For example, the boundary of a current $T$ is defined by $\p T(\o) = T(d\o).$  Stokes' theorem holds by definition.  Since $d^2 = 0,$ it follows that $\p^2 = 0$, yielding the current complex 
  $${\cal T}_{k+r}^0 \buildrel \p \over \to  \cdots \buildrel \p \over \to {\cal T}_{k+1}^{r-1}  \buildrel \p \over \to  {\cal T}_k^r \buildrel \p \over \to {\cal T}_{k-1}^{r+1} \buildrel \p \over \to \cdots \buildrel \p \over \to {\cal T}_{0}^{k+r}.$$

  Questions of regularity of currents can be a serious issue.  For example, does the boundary of a current correspond to the boundary defined geometrically? Is a current solution smooth?  What is its support?  These problems can be difficult to solve.   In a way, spaces of currents are too big, and we look for proper subspaces that are simpler to work with.  Examples include the normal and integral currents of Federer and Fleming \cite{ff} which have been important in geometric measure theory.   
  
We identify   proper subspaces of currents ${\cal J}_k^r \subset {\cal T}_k^r$, called {\itshape  $k$-chainlets of degree $r$}, that give us a  natural covariant {\itshape chainlet complex} for domains that is algebraically closed under the basic operators of calculus.  
 $${\cal J}_{k+r}^{0} \buildrel \p \over \to  \cdots \buildrel \p \over \to {\cal J}_{k+1}^{r-1}  \buildrel \p \over \to  {\cal J}_k^r \buildrel \p \over \to {\cal J}_{k-1}^{r+1} \buildrel \p \over \to \cdots \buildrel \p \over \to {\cal J}_{0}^{k+r}.$$  
While currents are dual to forms, chainlets are a {\itshape  predual} to forms.  More precisely, let ${\cal B}_k^r$ denote the space of  $k$-forms,   of class $C^{r-1+Lip}$, with a bound on each of the derivatives of order $s, 0 \le s \le r.$  (We do not require that the ambient space  be compact.) Then the space of bounded linear functionals on ${\cal J}_k^r$ is precisely ${\cal B}_k^r$.  
 Chainlets are not reflexive and thus ${\cal J}_k^r \subset ({\cal J}_k^r)^{**} = ({\cal B}_k^r)^* \subset {\cal T}_k^r$.  An example of a current that is not a chainlet in $V=\R^n$ is $\R^n$ itself.  However, even for compact ambient manifolds, chainlets form a proper subspace of currents since the space of chainlets is normed and the space of currents is not.

   Each differential $k$-form $\o \in {\cal B}_k^r$ is represented by a $k$-chainlet $J_{\o}$ in the sense that $\int_{J_{\o}} \eta = \int \o \wedge \star \eta$ for all $\eta \in {\cal B}_k^r.$  (The LHS is the {\itshape  chainlet integral} of \S\ref{chainletcomplex}. The RHS is the Riemann integral.)  Therefore, ${\cal B}_k^r$ is naturally immersed in ${\cal J}_k^r$. (See \cite{ravello}, \cite{gmt}.)
   $${\cal B}_k^r \subset {\cal J}_k^r \subset {\cal T}_k^r.$$    All three of these spaces are Banach spaces for each $0 \le r < \i$, and the inclusions are strict.  Each is dense in the larger spaces.   Define the direct limit ${\cal J}_k^{\i} = \bigcup {\cal J}_k^r$ and the inverse limit ${\cal B}_k^{\i} = \bigcap {\cal B}_k^r$. The space ${\cal B}_k^{\i}$ is a Frechet space, while ${\cal J}_k^{\i}$ is a normed space.  It is a direct limit of Banach spaces and contains a dense inner product space, while    currents ${\cal T}_k^{\i}$ are a topological vector space.

   The chainlet complex contains a dense subcomplex of discrete chains ${\cal P}_k^r(V)$,  called {\itshape  polypolar chains} with each  chain supported in finitely many points.  
   $${\cal P}_{k+r}^0 \buildrel \p \over \to  \cdots \buildrel \p \over \to {\cal P}_{k+1}^{r-1} \buildrel \p \over \to  {\cal P}_k^r \buildrel \p \over \to {\cal P}_{k-1}^{r+1} \buildrel \p \over \to \cdots \buildrel \p \over \to {\cal P}_0^{r+1}.$$   Wedge product and inner product are both  defined on this discrete subcomplex.  But there is no continuity of either of these products in the chainlet norms.  The chainlet complex has neither a wedge product nor an inner product since it includes polyhedral chains and $L^1$ functions.  (Polyhedral chains have no wedge product and $L^1$ functions have no inner product.)  But the space of chainlets contains a dense, inner product space --  the algebra of monopolar $k$-chains.

 \subsection*{Acknowledgements}
 I would like to thank Morris Hirsch for his continued encouragement,  support, and invaluable feedback for this work which evolved over a period of two decades.   He believed in the potential of this theory long before anyone else,   when there was only a scent in the air, and I seemed to be hacking through the thick undergrowth with a machete to find its source. 
   James Yorke has also given encouragement and helpful advice over the years for which I am grateful.   
   
    I wish to thank Alain Bossavit and Robert Kotiuga  who helped me appreciate the historical perspective and interests of theoretical and computational engineering.    I also thank the   students at the Ravello Summer School of Mathematical Physics, and those in my Spring 2005 graduate seminar, especially  Patrick Barrow and Alan Shinsato  for encouraging me towards the full development of this program.   Finally, I thank the wonderful undergraduates in my Berkeley 2006 experimental course on chainlets for their fresh energy and enthusiasm.  
    
  \newpage

   \section{The Koszul complex}\label{koszul}
  	\subsection{Exterior algebra}  Let $V$ be a vector space over a field $\F = \R$ or $\C$.  Let $T(V)$ denote its tensor algebra.  Let $I$ be the two sided ideal of $T(V)$, generated by all elements of the form $v \otimes v, v \in V$ and define $\Lambda(V)$ as the quotient $$\Lambda(V) = T(V)/I.$$   Use the symbol $\wedge$ for multiplication, or {\itshape \bfseries exterior product} in $\Lambda(V).$   Since $v \wedge v = 0$ we deduce $v \wedge w = - w \wedge u$ and $v_1 \wedge \dots \wedge v_k = 0$ whenever $W= (w_1, \dots, w_k)$ are linearly dependent in $V$.  Elements of the form  $v_1 \wedge \dots \wedge v_k$ are called  {\itshape \bfseries simple $k$-vectors}. We sometimes denote these by 
 $$ \a_k(W)  = w_1 \wedge \dots \wedge w_k$$ 
 where $W$ is the list $(w_1, \dots,w_k).$  For example, $\a_2(v,w) = v\wedge w.$  If $W_1 = (w_{1,1}, \dots, w_{1,k})$ and  $W_2 = (w_{2,1}, \dots, w_{2,l}),$ let   $$(W_1, W_2) = (w_{1,1}, \dots, w_{1,k}, w_{2,1}, \dots, w_{2,l}).$$  Thus $\a_{k}(W_1)\wedge \a_{l}(W_2) = \a_{k+l}(W_1, W_2).$
 
  The subspace of $\Lambda(V)$ generated by all simple $k$-vectors is known as the {\itshape \bfseries $k$-th exterior power} of $V$ and is denoted by $\Lambda_k(V)$. The exterior algebra can be written as the direct sum of each of the k-th powers:
   $$ \Lambda(V) = \bigoplus_{k=0}^{\infty} \Lambda_k V$$ noting that $\Lambda_0(V) = \F$ and $\Lambda_1(V) = V.$
The exterior product of a $j$-vector and an $k$-vector is a $(j+k)$-vector. Thus, the exterior algebra forms a graded algebra where the grade is given by $k$. 

  We may replace $v^2 = v \otimes v$ with any quadratic form $Q(v)$ and obtain geometric differentials of the Clifford algebra.  This extension will be treated    treated in a sequel.
   
 \subsection{Symmetric algebra}
The {\itshape \bfseries symmetric algebra} of $V$ is defined as follows.  Let $J$ be the two-sided ideal of $T(V)$ generated by all elements   of the form $v \otimes u - u \otimes v.$   Define $S(V) = T(V)/J.$  The {\itshape \bfseries symmetric product} of $u,v \in V$ is denoted $uv = vu$.  For $j \ge 1$, denote the $j$-th symmetric product  by $$\nabla^j_U = u_1 u_2 \cdots u_j$$ where $U$ is the list $(u_1, \dots, u_j).$   The subspace of $S(V)$ generated by the $j$-th symmetric products is denoted $S^j(V).$  There is a direct sum decomposition of S(V) as a graded algebra into summands  $S(V) = \oplus S^j(V)$ where  $S^0(V) = \F$ and $S^1(V)=V.$  The space $S^j(V)$ is the {\itshape \bfseries $j$-th symmetric power} of $V$.  Elements of $S^j(V)$ of the form $u_1u_2 \cdots u_j, j \ge 1$, are denoted   by  $\nabla^j_U$ where $U= (u_1, u_2, \dots, u_j).$  We observe that $\nabla^j_U$ is independent of the order of the vectors of $U$ and permits duplications, whereas $\a_k(W)$ depends on the order of $W$.  
 
Define the associative, unital algebra $$X(V) = \oplus_{j ,k} S^j(V) \otimes \Lambda_k(V).$$  Denote   $\nabla^j_U \a_k = \nabla^j_U \otimes \a_k,$ where $\nabla_U^j \in S^j(V)$ and $\a_k \in \L_k(V).$   
 The integer $k$ is the {\itshape \bfseries geometric dimension} of $\nabla^j_U \a(V^k)$, and $j$ is its {\itshape \bfseries order}. 
The product $\cdot$ in $X(V)$   is defined for simple vectors $ \a(W_1),  \a(W_1)$ by  
 $$\nabla^{j_1}_{U_1} \a(W_1) \cdot \nabla^{j_2}_{U_2} \a(W_2) = \nabla^{j_1+j_2}_{(U_1, U_2)} \a(W_1, W_2)/2^{j_1+j_2},$$   for $j_1, j_2 \ge 1,$ and extended by linearity.  For $j_1 = 0$ or $j_2 = 0,$ the product is merely scalar multiplication.  The operator $\nabla_U^j:X(V) \to X(V)$ is defined by $$\nabla_U^j (\nabla_{U'}^{j'} \a) = \nabla^{j+j'}_{(U, U')} \a.$$

\subsection{Mass and direction of  simple  $k$-vectors} We choose a preferred basis $(e_1, \dots, e_n)$ of $V$ and use it to define an inner product on $V$:  $<e_i, e_j> = \d_{ij}.$ 
   Define {\itshape \bfseries mass} of a simple $k$-vector by $$M(\a(w_1, \dots, w_k) ) = det(<w_i, w_j>).$$  If a different basis is chosen, the two masses will be proportional.  The resulting normed spaces will be the same, and the theory will be identical.   
 The {\itshape \bfseries orientation}  of the subspace $W$ of $V$ spanning the list $(w_1, \dots,w_k)$ is the orientation of the list $(w_1, \dots,w_k).$ (Two linearly independent lists have the same orientation if and only if the determinant of the change of basis matrix is positive.)  The {\itshape \bfseries $k$-direction} of a simple $k$-vector  $\a(w_1, \dots,w_k)$ is the oriented  $k$-dimensional subspace  spanning the list $(w_1, \dots,w_k)$.   The $k$-direction of a simple $k$-vector is independent of the preferred basis.

\begin{lemma}\label{dir}
Two simple $k$-vectors $\a$ and $\b$ are equal if and only if their masses and directions are the same.  
\end{lemma}

The proof follows directly from the multilinearity properties of tensor product.        
\subsection{  The Banach algebra $X(V)$}

Define $$\|\nabla_U^j \a\|_j =  |u_1| \cdots |u_j| M(\a)$$ where $U = (u_1, \dots, u_j).$  A basis of $V$ generates a basis of $S^j \otimes \L_k$.  Let  $A^j = \sum_i a_i \nabla_{U_i}^j \a_i$, written in  terms of this basis, and  define 
$$\|A^j\|_j = \sum |a_i|\left \| \nabla_{U_i}^j \a_i\right\|_j.$$  
Finally, for   $A = \sum_j A^j \in S(V)$, define $$\|A\| = \sum_j \|A^j\|_j.$$  
\begin{proposition}\label{jnorms}    $\|\cdot\|$ is a norm on the algebra$X(V)$ satisfying $$\|A \cdot B \|  \le \|A\|  \|B\| .$$ 
\end{proposition}

\begin{proof}  Supppose $A = \sum_{j=1}^s A^j$ and $\|A\| > 0.$   We may assume that the terms of $A^j =\sum_i a_i  \nabla_{U_i}^j \a_i$ are written in terms of a basis.     By the triangle inequality,  $\|A^j\|_j > 0$ for some $j$.  Hence   
$\| \nabla_{U_i}^j \a_i\|_j > 0$  for some $i$.   By definition of the norm,  and since mass is a norm, it follows that   $ \nabla_{U_i}^j \a_i \ne 0$.  Since $A^j$ is written in terms of a basis, we know $A^j \ne 0$, and thus $A \ne 0$ since we are using a direct sum.     The other properties of a norm follow easily from the definitions. 

The inequality is a consequence of the triangle inequality and 
$$\begin{aligned} \| \nabla_{U_i}^{j_i} \a_i \cdot  \nabla_{U_{\ell}}^{j_\ell} \a_{\ell}\|_j &= \|D^{j_i+ j_{\ell}}_{(U_i, U_{\ell})} \a_i \wedge \a_{\ell}\|_j \\&= |U_i||U_{\ell}| M(\a_i \wedge \a_{\ell})\\&\le  |U_i||U_{\ell}| M(\a_i)M( \a_{\ell}) \\&= \| \nabla_{U_i}^{j_i} \a_i\|_{j_i}\| \nabla_{U_{\ell}}^{j_{\ell}} \a_{\ell}\|_{j_{\ell}} .\end{aligned}$$
\end{proof}

 \subsection{The covariant complex}   
 \subsubsection*{Prederivatives $\nabla_u$}  Let $u \in V$ and $\nabla^j_U \a \in X(V).$  Define the {\itshape \bfseries prederivative (in the direction $u$)}  by
 $$\nabla_u:S^j \otimes \L_k \to S^{j+1} \otimes \L_{k}$$ by 
 $$\nabla_u(\nabla^j_U \a) = \nabla_{(u, U)}^{j+1} \a.$$  In the next section we give a geometric interpretation of this operator.  In a sequel, we show it is a derivation.
 
 We next define the boundary operator   $$\p:S^j \otimes \L_k \to S^{j+1} \otimes \L_{k-1}, k \ge 1.$$  Define
 $$ \p v := \nabla_v(1).$$   
   There is a unique extension of $\p$ to $X(V)$ making the boundary operator into a derivation.  In particular, $$\p(u \wedge v) = \p u \cdot v - u \cdot \p v.$$  
   In general,    $$\p:S^j \otimes \L_k \to S^{j+1} \otimes \L_{k-1}$$ is defined recursively:  
   
   Assume the boundary of a simple $k$-vector $\a$ has been defined, and  $v \in V$.  Define $$\p(\a \wedge v) = \p \a \cdot v + (-1)^{k} \a \cdot \p v \in S^1 \otimes \L_{k}.$$   For $j \ge 0$, define $$\p (\nabla_U^j \a) = \nabla_U^j(\p \a).$$

    We obtain a linear mapping $$\p: S^{j-1}(V) \otimes \L_k(V) \to S^j(V) \otimes \L_{k-1}(V)$$  which is a derivation.  It follows that $\p \circ \p = 0.$

We conclude that   $X(V)$ is a unital, associative, bigraded differential Banach algebra.  The product of $X(V)$ is associative, bilinear, graded commutative, and the boundary operator is a derivation:
If $A,B \in X(V)$, then  $$\p(A \cdot B) = (\p A) \cdot  B + (-1)^{dim( A)}  \cdot (\p B).$$   

For each $0 \le k+j = c, 0 \le k \le n$ and $j \ge 0$,  we have the bigraded, differential, covariant complex
$$ S^0(V) \otimes \L_{j+k}(V) \buildrel  \p \over \to S^1(V) \otimes \L_{j+k-1}(V) \buildrel  \p \over \to  \cdots  \buildrel  \p \over \to S^{j+k}(V)\otimes \L_0(V).$$

We next see how the boundary operator relates to the   prederivative operator in the case of a simple $k$-vector.

\begin{lemma}\label{sums} If  $\a$ is a simple $k$-vector, then $\p \a$ is the sum of $k$   prederivatives of simple $(k-1)$-vectors. 
\end{lemma}

\begin{proof}  The proof proceeds by induction on $k$.  The result holds by definition for $k = 1.$  Assume it holds for simple $(k-1)-$ vectors.
 If $\a$ is a simple $k$-vector then  $\a = \b \wedge v$  where $\b$ is a simple $(k-1)$-vector, then $\p \a = \p \b \cdot v + (-1)^{k-1} \b \cdot  \p v.$  By induction, $\p \b$ is the sum of $(k-1)$   prederivatives of simple $(k-1)$-vectors.  But $\b \cdot \p v = \nabla_v \b$ is also the   prederivative of a simple $(k-1)$-vector.  
\end{proof}

\subsection{The contravariant Koszul complex}     Define the exterior derivative operator $$d:  S^j(V) \otimes \L_k(V^*) \to S^{j-1}(V) \otimes \L_{k+1}(V^*)$$ as follows:  $d(v \otimes 1) = 1 \otimes v.$  Much as in the preceding section,  $d$ has a unique extension so it is a graded derivation on the product. 

For each $0 \le k+j = c, 0 \le k \le n$ and $j \ge 0$,  we have the {\itshape  contravariant Koszul complex}
$$    S^0(V) \otimes \L_{j+k}(V^*) 
\buildrel  d \over \leftarrow \cdots  \buildrel d \over  \leftarrow  S^{j+k-1}(V) \otimes \L_{1}(V^*)
 \buildrel  d \over \leftarrow  S^{j+k}(V) \otimes \L_{0}(V^*)  $$  

\begin{theorem}
$  S^{j}(V) \otimes \L_{k}(V^*) \subset  (S^{j}(V) \otimes \L_{k}(V))^*.$ 
\end{theorem}

Even though   $\L_k(V^*) \cong (\L_k(V))^*$ and $S^j(V^*) \cong (S^j(V))^*$ the tensor products of dual spaces is not isomorphic to the dual of tensor products.  The exterior derivative that we have defined as dual to the boundary operator is an operator on the space $(S^{j}(V) \otimes \L_{k}(V))^*$, and thus is also an operator on $ S^{j}(V) \otimes \L_{k}(V^*).$   On the other hand, the exterior derivative defined in the contravariant Koszul complex $ S^{j}(V) \otimes \L_{k}(V^*) $ by ``moving covectors from the symmetric side to the antisymmetric side'' is not naturally extendable to  $(S^{j}(V) \otimes \L_{k}(V))^*.$

 The algebra of polynomials is isomorphic to the symmetric algebra and can be useful as coefficients of $k$-covectors in $S^{j}(V) \otimes \L_{k}(V^*).$ Again, this does not have natural extension into $(S^{j}(V) \otimes \L_{k}(V))^*.$   For  $(S^{j}(V) \otimes \L_{k}(V))^*$ it is more natural to use differential forms with coefficient functions with bounded derivatives, such as trigonometric functions or  smooth functions with compact support. This touches on a philosophical difference between Taylor's theorem approximations by polynomials, vs approximations with Fourier series or wavelets with their superior convergence rates.  
 
 In summary, the covariant complex $X(V)$ leads to a more general theory than the contravariant complex $X(V^*)$.   It is simpler to work with conceptually as it is based  upon physical concepts.\footnote{There is a foundational difference between the approach of \cite{arnold} and our own.  By building upon the {\itshape prederivative} operator $\nabla_u$ of the Koszul chain complex, rather than the {\itshape exterior derivative} operator of the Koszul cochain complex, we generate a discrete theory with a geometrical basis rather than a basis of differential forms.
} The basic operators of pushforward, boundary, and perp are more natural than the duals of pullback, exterior derivative, and Hodge star.

 	\section{The polypolar chain complex}\label{pointed}  
	\subsection{The algebra of polypolar chains} Consider  the product space $V \times X(V).$     A {\itshape \bfseries $j$-polar $k$-vector in $V$ with support at $p$} is a pair $(p; \nabla_U^j\a) \in V \times \L_k^j(V).$   We form the vector space ${\cal P}_k^j(V)$ of  {\itshape \bfseries $j$-polar $k$-chains }  of formal sums  $A = \sum_{i=1}^s (p_i; \nabla_{U_i}^{j_i}\a_i)$ subject to the relation $(p; \nabla_U^j\a) = (p;-\nabla_U^j\b)$ if $\a$ and $\b$ have the same mass and $k$-direction, but with opposite orientation.  Furthermore, $(p; \nabla_U^j\a) + (p; \nabla_{U'}^{j'} \a') = (p;  \nabla_U^j\a +\nabla_{U'}^{j'} \a'),$   and $t(p;\nabla_U^j\a) = (p; t\nabla_U^j\a), t \in \F.$       Define ${\cal P}_k = \oplus_j {\cal P}_k^j$ and ${\cal P} = \oplus_k {\cal P}_k.$   For $j = 0, 1, 2$, we call $j$-polar $k$-chains \emph{monopolar, dipolar} and \emph{quadrupolar} $k$-chains, respectively.
	We are especially interested monopolar $k$-chains as they are dense in spaces of chainlets.  For simplicity, consider monopolar $1$-vectors. The space ${\cal P}_1^0(V)$ differs from the affine space of $V$ in the following important way.  The space of monopolar $1$-chains is a vector space, equipped with the translation operator $T_v(p;\nabla_u \a) = (p+v; \nabla_u \a).$  The affine space associated to a vector space $V$ is not a vector space, as there is no addition, only subtraction $p-q.$     A monopolar $k$-vector $(p; \a).$ separates the two basic roles of a vector into its two parts.  In the first slot of $(p; \a)$, we think of the vector $p \in V$ as a point.   It is merely the support, or location, of the $k$-vector $\a$ in the second slot.  The pair becomes a ``smart point'' which contains much more information than just $p$ alone.  	
	  We make use of two projections: $$supp:{\cal P}_k^j(V) \to V,$$ mapping $P = \sum (p_i; \nabla_{U_i}^{j_i}\a_i)$ to the point set $supp(P) = \cup p_i$ and $$Vec_k^j:{\cal P}_k^j(V) \to \L_k^j(V)$$ mapping $P = \sum (p_i; \nabla_{U_i}^{j_i}\a_i)$ to $\sum  \nabla_{U_i}^{j_i}\a_i.$

	\paragraph{\itshape \bfseries Wedge product on ${\cal P}$} Define $$(p; \a) \wedge (q; \b) = \begin{cases}0, & p \ne q\\ (p; \a \wedge \b), & p = q. \end{cases}.$$  For 
$A = \sum (p_i;A_i)$ and $B = \sum  (q_j; B_j) \in V \times X(V),$ define $$A \wedge B = \sum   (p_i; A_i \cdot B_j)$$ where $p_i = q_j.$  This reduces to wedge product of $k$-vectors if   $A$ and $B$ have order zero.   For $A = \sum (p_i;A_i)$, define the norm $$\|A\| = \sum \|A_i\|.$$  Then $$\| A \wedge B\| \le \|A\|\|B\|,$$ 
making the space of monopolar chains ${\cal P}$ into an associative, unital, Banach algebra.  

  Define the {\itshape \bfseries $1$-difference monopolar $k$-vector} by $$\Delta^1_u(p;\a) =  T_u(p;\a) - (p;\a)$$  and the {\itshape \bfseries $j$-difference monopolar $k$-vector} recursively by $$\Delta_U^j(p;\a) = \Delta^1_{u}(\Delta^{j-1}_{U'}(p;\a)).$$ 
	where $U = (u, U').$   
	Define $\o(\Delta_u^1(p; \a))= \o(T_u(p; \a)) - \o(p; \a)$ and extend recursively to $\o(\Delta_U^j(p;\a))$ and by linearity to $\o(P), P \in {\cal P}_k^j.$

We next see how to develop calculus at a point by way of polyvectors.  
Define $$ \|\Delta^0(p; \a)\|_0 = M(\a),$$ and, for $j \ge 1$, define
  $\|\Delta^j_U(p; \a)\|_j = |U| M(\a)$ where $|U|   = |u_1| \cdots |u_{j}|, U=(u_1, \cdots, u_j).$
 
 Define ${\cal D}_k^j(V)$ to be the subspace of ${\cal P}_k^0(V)$ generated by $j$-difference $k$-chains $$D^j = \sum a_i \Delta_{U_i}^j(p_i; \a_i).$$

	\subsection{Differential forms on monopolar chains}  
	  A {\itshape  differential $k$-form} $\o$ is defined to be a linear functional on the space of monopolar $k$-chains ${\cal P}_k(V).$   	  The {\itshape  support} of a $k$-form is a closed set $K$ defined as follows:   A point $x$ is in the complement of $K$ if there exists a neighborhood $U$ of $x$ and missing $K$, such that $\o(x;\a) = 0$ for all pairs $(x;\a), x \in U, \a \in \L_k(V)$.

	   Denote the operator norm $$|\o|^{\natural_{0}}= \|\o\|_{0} = \sup\left\{ |\o(x;\a)| : M(\a) = 1  \right\}.$$  We say that $\o$ is {\itshape  bounded measurable} if   $|\o|^{\natural_{0}} < \i.$  The space of all bounded measurable forms is denoted $ {\cal P}_k(V)^\star$

   Let $\o \in  {\cal P}_k(V)^\star$ and $j \ge 0.$  
	 Define
	 $$\|\o\|_{j} = \sup\left\{|\o(\Delta_U^j(x;\a))|:    \|\Delta_U^j(x;\a) \|_j = 1  \right\}.$$    
	 Define the norm
	 \[|\o|^{\natural_{r}} = \max\{ \|\o\|_{0}, \dots, \|\o\|_{r}\}\]

	 \begin{proposition}
$|\o|^{\natural_{r}} $ is a norm on  the subspace of bounded Lipschitz forms  with $|\o|^{\natural_{r}} < \i$.
\end{proposition}

   The completion of the space of bounded measurable $k$-forms  with  $|\o|^{\natural_{r}} < \i$ is denoted ${\cal B}_k^r(V).$\footnote{In coordinates, this space is equivalent to the space of $k$-forms with coefficients of class $C^{r-1+Lip}$, i.e.,   the   Lipschitz constant of each derivative of order $(r-1$) is uniformly bounded in $V$. }   Examples include $sin(2x) dx$, $f(x) dx$ where $f$ has compact support.
 Nonexamples include $xdx$ since $\|xdx\|_0 = \infty.$

	\subsection{Wedge product of forms}       Given $\o \in {\cal B}_k^r$ and $\eta \in {\cal B}_m^s$, define
  
  $$\begin{aligned} \o \wedge &\eta(p;\a(v_1, \dots, v_{k+m}))  \\&=   \sum_{\s \in S_k}\o(p;\a(v_{\s(1)}, \dots, v_{\s(k)}) \eta( \a(p; v_{\s(k+1)}, \dots, v_{\s(k+m)}).\end{aligned}$$

\begin{proposition} Suppose $\o \in {\cal B}_k^r, \eta \in {\cal B}_m^s$.  Then 
$$|\o \wedge \eta|^{\natural_{r+s}} \le |\o|^{\natural_{r}} |\eta|^{\natural_{s}}.$$
\end{proposition} 

\begin{proof}
We prove that $$\|\o \wedge \eta\|_{t+u} \le \|\o\|_t \|\eta\|_u$$ for all $0 \le t \le r, 0 \le u \le s.$
The LHS is the supremum of terms of the form
$\frac{(\o \wedge \eta)(\Delta_U^{t+u} \a_{k+m})}{\|\Delta_U^{t+u} \a_{k+m}\|_{t+u}}.$ We may assume wlog that  $\a_{k+m} = \a(e_1, \dots, e_{k+m})$ where $(e_1, \dots, e_{k+m})$ is orthonormal.  Then $M(\a_{k+m}) = M(\a_k)M(\a_m)$ for any permutation of the vectors of $\a_{k+m}.$  Split the list $U^{t+u} $ into two lists $U^{t+u} =(U_1^t, U_2^u)$.

The result now follows since wedge product is averaged over all such permutations.      
\end{proof}

	\subsection{Norms on ${\cal P}_k^0(V)$}   Denote the dual norm on ${\cal P}_k^0(V)$ by $$|P|^{\natural_{r}} = \sup_{\o \in {\cal B}_k^r} \frac{ \o(P)}{|\o|^{\natural_{r}}}, P \in {\cal P}_k^0(V).$$  

By duality, $|P|^{\natural_{r}}$ is a norm on the space ${\cal P}_k^0(V).$

For $D^j \in  {\cal D}_k^j(V)$, define $$\|D^j\|_j = \sup \frac{\o(D^j)}{\|\o\|_j}.$$ The next lemma follows directly from the definitions.
\begin{lemma}
$|D^j|^{\natural_{j}} \le \|D^j\|_j.$
\end{lemma}
\begin{theorem}\label{method} If $P \in {\cal P}_k^0$ and $ r \ge 0$, then
$$|P|^{\natural_{r}} = \inf\{\Sigma_{j=0}^r \|D^j\|_j: P = \Sigma_{j=0}^r D^j, D^j \in {\cal D}_k^j\}.$$
\end{theorem}

\begin{proof}  Suppose $P = \sum D^j.$  By the triangle inequality
$$|P|^{\natural_{r}} \le \sum |D^j|^{\natural_{r}} \le \sum \|D^j\|_i.$$ 

  On the other hand, let $\e > 0$.  There exists $P = \sum D^j$ such that $RHS > \sum \|D^j\|_j  - \e.$   Then
$$ \begin{aligned} |P|^{\natural_{r}} = \sup_{\o \in {\cal B}_k^r} \frac{\o(P)}{|\o|^{\natural_{r}}} &=  \sup_{\o \in {\cal B}_k^r} \frac{\o(D^0) + \cdots +\o(D^r)}{\max\{\|\o\|_0, \dots, \|\o\|_r\}} \\&\le \sup_{\o \in {\cal B}_k^r}\left\{
\frac{\o(D^0)}{\|\o\|_0} + \cdots + \frac{\o(D^r)}{\|\o\|_r}\right\}  
  \\& \le  \sum_{j=0}^r \|D^j\|_j  < RHS + \e.\end{aligned}$$  Since this holds for all $\e > 0$, the result follows.
\end{proof}

\subsection{Prederivative operator $\nabla_u$}        
\begin{lemma}\label{mainlimit}  If $\o \in {\cal B}_k^2$, $p \in V$, and $\a \in \L_k(V)$, then
 $$\lim_{t \to 0} \left\{
 \o\left(p;\frac{\Delta_{tu}(p; \a)}{t}\right)
 \right\}$$ exists and is unique.  \end{lemma}

 The proof to this appears in \cite{ravello}.  If $\o$ is only Lipschitz, Radamacher's theorem shows this limit exists a.e.  If $\o \in {\cal B}^{1+\e}$, the limit exists and is continuous.  

Define the directional derivative $$D_u \o (p;\a)   = \lim_{t \to \i} \o\left(p; \frac{\Delta_{tu}(p; \a)}{t}\right),$$  and $$\o(p; \nabla_u \a) = D_u \o (p;\a).$$

\begin{lemma}\label{Du}  $D_u$ is bilinear in $u$ and $\a$ and satisfies
$$|D_u \o|^{\natural_{r-1}} \le |u||\o|^{\natural_{r}}.$$

\end{lemma}

\begin{proof} It suffices to show $$\|D_u \o\|_{j-1} \le |u| \|\o\|_{j}.$$
But $$ \|D_u \o\|_{j-1}  = \sup\frac{(D_u \o)(\Delta_U^{j-1} (p;\a))}{\|\Delta_U^{j-1} (p;\a)\|_{j-1}}   =
\sup \lim_{t \to 0} \frac{\o(\Delta^j_{(u,U)}(p; \a) |u|}{t\|\Delta^j_{(u,U)}(p;\a)\|_j}
 \le |u| \|\o\|_{j} $$  
\end{proof}

\begin{lemma}\label{contnab}
$|\nabla_u P|^{\natural_{r}} \le |u||P|^{\natural_{r-1}}.$
\end{lemma}

\begin{proof}  By Lemma \ref{Du}
$$\begin{aligned}  |\nabla_u P|^{\natural_{r}} = \sup \frac{\o(\nabla_u P)}{|\o|^{\natural_{r}}}  =  \sup \frac{D_u \o(P)}{|\o|^{\natural_{r}}} 
\le \sup \frac{|D_u \o|^{\natural_{r-1}}|P|^{\natural_{r-1}}}{|\o|^{\natural_{r}}} \le |u||P|^{\natural_{r-1}}. \end{aligned}$$
\end{proof}

\subsection{Exterior derivative $d$}	 
  By Lemma \ref{sums} the boundary of a simple $k$-vector $\a$ is the sum of $k$   prederivatives of simple $(k-1)$-vectors $\b_i$. That is, $\p \a = \sum \nabla_{u_i}\b_i.$  It is often convenient to assume that the directions of translation are orthogonal to the $\b_i$.  This follows if $\a = \a(v_1, \dots, v_k)$ where the list $(v_1, \dots , v_k)$ is orthogonal.  But $\a$ always has such a representative by Lemma \ref{dir}.
 
 Define the exterior derivative on forms by $$d\o(p; \a) = \o(p; \p \a).$$
  
\begin{lemma}\label{phi} Let $\o \in {\cal B}_k^2$ and $\phi \in {\cal B}_k^2$.  Then 
\begin{enumerate} 
\item $d \phi(p; u) =  \nabla_u \phi(p)$
\item $d\o(p; \a) = \sum \nabla_{u_i}\o(p; \b_i).$
\end{enumerate} 
\end{lemma}

\begin{proof} Part (i) follows since
 $$d \phi(p; u)  = \phi(p; \p u) = \phi(p; \nabla_u(1)) = D_u \phi(p).$$  For Part (ii), we have $$d\o(p; \a) = \o(p; \p \a) =   \o(p; \Sigma \nabla_{u_i} \b_i )= \sum D_{u_i} \o(p; \b_i).$$
\end{proof}

\begin{proposition}
\begin{enumerate}
\item $d(\o + \eta) = d\o + d \eta$
\item $d(\o \wedge \eta) = d \o \wedge \eta + (-1)^k \o \wedge d \eta$
\item $d(d \o) = 0$
\item $d\phi = \sum D_{e_i} \phi de_i$ if $(e_1, \dots, e_n)$ is orthonormal.

\end{enumerate}

\end{proposition}
\begin{proof}
Part (i) follows from linearity of boundary operator.
Parts (ii) is a consequence of Lemma \ref{phi} and the definition of wedge product.  (See \cite{flanders}, for example.)  (iii) follows since $\p \circ \p = 0.$  

Proof of (iv): $$d \phi(p; \sum a_i e_i) = \sum d \phi(p; e_i) = \sum D_{e_i}\phi(p) = \sum D_{e_i} \phi de_i(p; e_i).$$
\end{proof}

\begin{proposition}\label{dbound} If $\o \in {\cal B}_k^r$, then
$$|d\o|^{\natural_{r-1}} \le (k+1) |\o|^{\natural_{r}}.$$
\end{proposition}   

\begin{proof}   Let $\a$ be a simple $(k+1)$-vector. Suppose $\p \a = \sum \nabla_{u_i} \b_i$ where $u_i$ is orthogonal to the $k$-direction of $\b_i$.  Therefore,  
$$\begin{aligned} \|d\o\|_{j-1}  
= \sup\frac{d\o(\Delta_U^{j-1} \a)}{\|\Delta_U^{j-1} \a\|_{j-1}} &= 
\sup\frac{\o(\Delta_U^{j-1} \p \a)}{\|\Delta_U^{j-1} \a\|_{j-1}}
\\&\le  
 \sup \frac{\sum \o(\Delta^j_{(u,U)}\b_i)}{\|\Delta_U^{j-1} \a\|_{j-1}} \\&
 \le (k+1) \frac{\o(\Delta^j_{(u,U)} \b_i)}{\|\Delta^{j-1}_U \b_i \wedge u_i\|_{j-1}} \\& \le (k+1) \|\o\|_j \frac{|u_i||U| M(\b_i)}{|U|M(\b_i \wedge u_i)} \\&= (k+1) \|\o\|_j.
 \end{aligned}$$
\end{proof}

We may now extend the definition of forms recursively so they are defined on each ${\cal P}_k^r.$    Assume that forms of class ${\cal B}_k^{j+1}$ have been defined on  ${\cal P}_k^j, j < r.$  If  $\o \in {\cal B}_k^{r+1}$, then $D_u \o \in {\cal B}_k^r.$  Define $$\o(p; \nabla_{u,U}^{r} \a) = D_u \o (p; \nabla_U^{r-1} \a).$$ We may now define
$$d \o(p; \nabla_U^r \a) = \o(p; \nabla_U^r \p \a).$$

The boundary operator extends by linearity to monopolar chains $P$.
Stokes' theorem for monopolar chains follows  $$ d\o(P) = \o(\p P).$$

  \begin{proposition}\label{bound} If $P \in {\cal P}_k^r$, then 
$|\p P|^{\natural_{r+1}} \le k|P|^{\natural_{r}}.$
\end{proposition}

\begin{proof} By Proposition \ref{dbound} 
$$|\p P|^{\natural_{r+1}}  = \sup \frac{\o(\p P)}{|\o|^{\natural_{r+1}}} = 
\sup \frac{d \o( P)}{|\o|^{\natural_{r+1}}}  \le k \sup \frac{|d \o|^{\natural_{r}}|P|^{\natural_{r}}}{|d \o|^{\natural_{r}}} = k|P|^{\natural_{r}}.$$
 \end{proof}
 
 By linearity $\p \circ \p = 0$, giving us a chain complex for each $k+r = c:$ 
 $${\cal P}_{k+r}^{0} \buildrel \p \over \to  \cdots \buildrel \p \over \to {\cal P}_{k+1}^{r-1}  \buildrel \p \over \to  {\cal P}_k^r \buildrel \p \over \to {\cal P}_{k-1}^{r+1} \buildrel \p \over \to \cdots \buildrel \p \over \to {\cal P}_{0}^{k+r}. $$

  \section{Smooth Pushforward}
  We first give a coordinate free definition of the $r$-norm on smooth mappings, and then relate this to a more familiar norm relying on coordinates. 
 
\begin{definition}
The $r$-norm of a mapping $f:V \to \R^n$ is defined by
\[\|f\|_{[j]} = \sup_{U \ne 0, \alpha \ne 0, x} \frac{|f_*\Delta_U^j(x;\alpha)|^{\natural_{j}}}{\|\Delta_U^j(x;\alpha)\|_j}\] and
\[|f|_{[r]} = \max\{\|f\|_{[0]}, \dots, \|f\|_{[r]}\}\]  We say $f \in {\cal B}^r$ if $|f|_{[r]} < \infty.$
\end{definition}  

  We remark that $\|f\|_{[1]} < C$ is equivalent to assuming that $f$ satisfies a Lipschitz condition. 
\[\frac{|f(x+u) -f(x)|}{|u|} < C\]

Let $Q_K(0)$ denote the $n-cube$ centered at the origin with edge length $2K.$
\begin{lemma} \label{mappingnorm}
Let $f = (f_1, \dots, f_n):V \to Q_K(0)$ in standard coordinates of $\R^n$ and $r \ge 0$.   Then for each $1 \le i \le n$, 
\[ |f_i|^{\natural_{r}}  \le K |f|_r.\]  
\end{lemma}

\begin{proof}
It suffices to show   each
\[   \|f_i\|_j\le \|f\|_{[j]}\]

Let \[\omega_i(x;1) = \begin{cases} x_i, \quad &|x_i| \le K\\K, \quad &|x_i| \ge K.\end{cases}\]
  Then 
\[\|\omega_i\|_0 \le K, \|\omega\|_s \le 1 \implies |\omega_i|^{\natural_{j}} \le K.\]
But 
\[\begin{aligned} |f_i(\Delta_U^j(x;1))| = \left| \int_{\Delta_U^j(x;1)} f_i\right| &= \left| \int_{f_*\Delta_U^j(x;1)} \omega_i\right| \\&\le |\omega_i|^{\natural_{j}}|f_*\Delta_U^j(x;1)|^{\natural_{j}} \\&\le K|f_*\Delta_U^j(x;1)|^{\natural_{j}} \end{aligned} \]
Therefore, \[\|f_i\|_j \le K \|f\|_{[j]} \implies |f_i|^{\natural_{r}} \le K|f|_r.\]
\end{proof}

 \begin{lemma}  Let $f \in 
  \mathcal{B}^2, u \in V, x \in V.$  Then each directional derivative 
  $D_u f_i(x;1)$ exists and is a $0$-form of class $\mathcal{B}^1$.  
\end{lemma}
  
\begin{proof} 
  \[D_u f_i(x;1) = f_i(x; \nabla_u(1)) = \lim_{t \to 0} f_i(\Delta_{tu}(x; 1/t))\] by  the main lemma.  
\end{proof}    

For $u = e_j$, then $D_u f_i(x;1) = \frac{\partial f_i}{\partial x_j}.$

\begin{definition}
   Let $f = (f_1,\ldots,f_n)$, $x \in \mathcal{V}$, and $u \in V$. Define the \emph{pushforward}  $f_*(x;u)$ by
   \[ f_*(x; u) = (f(x); D_u f_1(x), \cdots, D_uf_n(x))\]
\end{definition}

One way to interpret pushforwards is as follows. Let $f: \R^m \to \R^n$ be a smooth mapping. Then we can think of $f_{*_x}: \R^m \to \R^m$ as the best {\it linear} approximation of $f$ at $x$. 
Suppose $u \in \R^m$.  
Let $\sigma_u = \{x + tu: 0 \le t \le 1\}$ be the $1$-cell of $(x;u)$.  Then $f_*\sigma_u$ is a singular $1$-cell with endpoint $f(x)$.   The tangent vector to   $f_*\sigma_u$ at $f(x)$ is $f_*(x;u).$  Now $f_{*x}$ is linear and in coordinates  takes the form of the classical Jacobian matrix
\[\left(\begin{array}{cccc}\frac{\partial f_1}{\partial x_1}(x) & \frac{\partial f_1}{\partial x_2}(x) & \cdots & \frac{\partial f_1}{\partial x_m}(x) \\\frac{\partial f_2}{\partial x_1}(x) & \ddots  &   & \vdots \\\vdots &   &\ddots   &   \\ \frac{\partial f_n}{\partial x_1}(x) & &  \cdots & \frac{\partial f_n}{\partial x_m}(x)\end{array}\right)\]
 
   In order to extend to monopolar $k$-chains, we define
   \[ f_*(x ; u_1 \wedge \cdots \wedge u_k) = (f(x) ; f_* u_1 \wedge \cdots \wedge f_* u_k). \]   Define 
   \[ f_*(x ; \nabla_u \alpha) = (f(x) ; \nabla_{f_* u} f_* \alpha) \]

From here we provide the following useful lemma that relates pushforward to all the other operators we have studied:

\begin{lemma}\label{push}
\quad
\begin{enumerate}
   \item $f_* \nabla_u = \nabla_{f^* u} f_* $
   \item $(f \circ g)_* = g_* \circ f_*$
   \item $f_* \partial = \partial f_*$
\end{enumerate}
\end{lemma}

Note however that the pushforward and perp operators do not commute. A counterexample is the shear operator.

Another property that is standard is the chain rule. This follows since it holds for linear mappings.  

\begin{theorem}[Chain Rule]  Suppose $f:\R^m \to \R^n$ and $g:\R^n \to \R^p$ are smooth mappings.  Then 
\[(f \circ g)_*= f_*g_*.\]

\end{theorem}

\begin{proof}
$$ \begin{aligned} f_*g_*(x;v) &= f_*(g(x);g_{*x}v)   = (f(g(x)); f_{*g(x)}g_{*x}v) \\&= (f \circ g(x); (f\circ g)_{*x}v) =  (f\circ g)_*(x; v)\end{aligned}$$
\end{proof}
 
\begin{definition}  Let $f: \R^n \to R^p$ be a smooth mapping and $\omega$ a $k$-form on $\R^p$.  Define $f^* \omega$ on $\R^n$ by 
   \[ f^*\omega(x ; \alpha) = \omega f_*(x ; \alpha) = \omega(f(x) ; f_*\alpha) \]
\end{definition}

The change of variables theorem is immediate for monopolar chains.
\begin{theorem}
   Change of variables:
   \[ \int_P f^* \omega = \int_{f_* P} \omega \]
\end{theorem}

The pullback behaves similarly to the pushforward.

\begin{lemma}
\quad
\begin{enumerate}
   \item $f^*(\omega \wedge \eta) = f^* \omega \wedge f^* \eta$
   \item $(f \circ g)^* = g^* \circ f^*$
   \item $f^* d = d f^*$
\end{enumerate}
\end{lemma}

\begin{proof}
1. follows since it holds for covectors.  For 2. we apply the chain rule.  3.  follows from Lemma \ref{push}.
\end{proof}

Now we establish the continuity of the pushforward operator.  
\begin{theorem}  
   $|f^* \omega|^{\natural_r} \le |f|_{[r]} |\omega|^{\natural_r}$
\end{theorem}
\begin{proof}
\[\begin{aligned}
\|f^* \omega\|_j = \sup \frac{f^* \omega(\Delta_U^j(x; \alpha))}{\|\Delta_U^j(x;\alpha)\|_j} &\le \sup\frac{\omega(f_*\Delta_U^j(x; \alpha))}{\|\Delta_U^j(x;\alpha)\|_j}  \\&\le   |\omega|^{\natural_{j}} \sup  \frac{|f_*\Delta_U^j(x; \alpha)|^{\natural_{j}}}{\|\Delta_U^j(x; \alpha)\|_j} \\&\le \|f\|_{[j]} |\omega|^{\natural_{j}} \\&\le |f|_{[r]} |\omega|^{\natural_{r}}
\end{aligned}\]  The result follows.
\end{proof}
Now we can establish the continuity of the pushforward.
\begin{corollary} If $P$ is a monopolar chain and $f$ is a smooth mapping of class ${\cal B}^r$, then
  $|f_*P|^{\natural_r} \le |f|_{[r]}|P|^{\natural_r}$
\end{corollary}
\begin{proof}
\begin{align*}
      |f_*P|^{\natural_r} = \sup \frac{|\omega(f_* P)|}{|\omega|^{\natural_r}} 
       &= \sup \frac{|f^*\omega(P)|}{|\omega|^{\natural_r}}  
      \le \sup \frac{|f^*\omega|^{\natural_r}|P|^{\natural_r}}{|\omega|^{\natural_r}} 
       \\&\le \sup \frac{|f|_{[r]}|\omega|^{\natural_r}|P|^{\natural_r}}{|\omega|^{\natural_r}}
       = |f|_{[r]} |P|^{\natural_r}.
\end{align*}
\end{proof}

  \section{The operator $\perp$}

  Let $\a$ be a simple $k$-vector.  Define $\perp \a$ to be the simple $(n-k)$-vector with the same mass as $\a$ and with $(n-k)$-direction orthogonal to the $k$-direction of $\a$.   Define $$\perp (\nabla_u \a) = \nabla_u (\perp \a).$$  Extend by linearity to define $$\perp:{\cal P}_k^j \to {\cal P}_{n-k}^j.$$  (See   \cite{hodge} for details of the perp operator in $\R^n$.)
 
 Define the unit volume $n$-form  $dV$ by $$dV(\a) = \perp \a$$ for simple $n$-vectors $\a.$  By linearity $\int_P dV = \perp P$ for $P \in {\cal P}_n^0.$ 

 Define the unit volume $n$-vector by $vol = \a(e_1, \dots, e_n)$ where $(e_1, \dots, e_n)$ is an orthonormal basis of $V$. 
\begin{lemma}
\begin{enumerate}
\item $\perp \perp \a = (-1)^{k(n-k)} \a$
\item $\a \wedge \perp \a = M(\a)^2 vol$
\item $\phi \perp = \perp \phi.$
\end{enumerate}
\end{lemma}
 
\begin{proposition}[Continuity of the perp operator]  If $P  \in {\cal P}_k^r$, then 
$$|\perp P|^{\natural_{r}} = |P|^{\natural_{r}}.$$ 
\end{proposition}

Define $\p^* = \diamond = \perp \p \perp$ and  the geometric Laplace operator by 
$\square = \p \diamond+ \diamond \p.$ 
Define the Hodge star operator by $$\star \o(p; \a) = \o(p; \perp \a).$$ 
\begin{lemma}
$$|\star \o|^{\natural_{r}} = |\o|^{\natural_{r}}.$$
\end{lemma}  

\begin{proof}
$$\begin{aligned} \|\star \o\|_j = \sup \frac{\star \o(\Delta_U^j(p; \a))}{\|\Delta_U^j(p; \a)\|_j} &=  \sup \frac{ \o(\Delta_U^j(p; \perp \a))}{\|\Delta_U^j(p; \a)\|_j} \\&\le \frac{\|\o\|_j \| \Delta_U^j(p; \perp \a)\|_j}{\| \Delta_U^j(p;   \a)\|_j} \\&=  \|\o\|_j\end{aligned}.$$
\end{proof}

\begin{theorem}[Star theorem]  If $\o \in {\cal B}_k^r$ and $P \in {\cal P}_{n-k}^r$, then
$$\int_P \star \o = \int_{\perp P} \o.$$
\end{theorem}

 The Laplace operator on forms is given by
 $\Delta = d \d + \d d$ where $\d = \star d \star.$
 
\begin{proposition}
$\Delta \o ( P) =\p(\square P).$
\end{proposition}
\begin{theorem}[Star theorem]
$$\int_P \star \o = \int_{\perp P} \o.$$
\end{theorem}

\section{The chainlet complex}\label{chainletcomplex}
	   
Denote the completion of the space of the space ${\cal P}_k^0(V)$ under the  norm  $|P|^{\natural_{r}}$ by ${\cal N}_k^r(V)$ and call its elements {\itshape \bfseries $r$-natural $k$-chainlets in $V$}.

Compare this to the definition of currents $T_k^r$ with norm  
$$|T|^{\natural_{r}} = \sup_{\o \in {\cal B}_k^r} \frac{T(\o)}{|\o|^{\natural_{r}}}.$$ 
Currents are dual to differential forms.  That is, $({\cal B}_k^r)^* = T_k^r$.    Chainlets of ${\cal N}_k^r$ are a predual to differential forms in that $({\cal N}_k^r)^* = {\cal B}_k^r$.   The spaces are not reflexive since ${\cal N}_k^r$ is separable.  It contains a countable dense subspace whereas, ${\cal B}_k^r$ is not separable. (See Whitney \cite{whitney}.)  Therefore, $${\cal N}_k^r \subset ({\cal N}_k^r)^{**}= T_k^r.$$  
 
 Let $J\in {\cal N}_k^r$. Then there exists a sequence of monopolar chians $P_i \to J$ in the  $r$-norm.  Since the norms are decreasing, it follows that $P_i$ converges to a unique chainlet $f_{rs}(J)   \in {\cal N}_k^{r+s}, s \ge 0.$
 
\begin{lemma}
  The mapping    $$f_{rs}:{\cal N}_k^r \to {\cal N}_k^{r+s}$$     is a homomorphism.   Furthermore, $f_{rr}(J) = J$ for all $J \in {\cal N}_k^r$ and $f_{rt} = f_{st} \circ f_{rs}$, for all $r \le s \le t.$
\end{lemma}

Define  the direct limit ${\cal N}_k^{\infty}(V) = \cup {\cal N}_k^r(V).$

\begin{theorem}
${\cal N}_k^{\infty}$ is a vector space with norm $$|P|^{\natural_{\infty}} = \lim |P|^{\natural_{r}}.$$
\end{theorem}

\begin{proof}
 Let $P = \sum_{i = 0}^s (p_i; \alpha_i)$ be a monopolar $k$-chain that is not zero.  Then $(p_0; \alpha_0) \ne 0,$  say.  We may assume without loss of generality that the $k$-direction $L$ of $\alpha_0$ is the subspace spanned by  $(e_1, \dots, e_k)$.    Let $\eta = dx_1 \cdots dx_k$.     Now $\eta(P) =  \sum_{i = 0}^s \eta (p_i; \alpha_i) =   \sum_{i = 0}^s \eta(p_i; \beta_i)$ where $\beta_i$ is the projection of $\alpha_i$ onto $L.$  Since $\beta_i$ is top dimensional, there exist scalars $f(p_i)$ such that $\beta_i = f(p_i)e_1 \wedge  \cdots \wedge e_k$.  There exists a polynomial $Q$ defined on $L$ such that $\sum Q(p_i) f(p_i) > 0.$  Define $\o = Q \eta.$  Then $$\o(P) = Q \eta (P) = \sum Q(p_i) f(p_i) > 0.$$ Now each derivative of $\o$ is bounded above by the derivatives of $Q$.  It follows that  $|P|^{\natural_{\infty}}  > 0.$

 The other properties of the norm are straightforward to verify.  
\end{proof} 

Every monopolar $k$-cell $(p; \a)$ is naturally included in ${\cal N}_k^r$ for each $r$.  The pair is called a   {\itshape \bfseries $(0,k)$-pole}.  This corresponds naturally to an electric monopole for $k = 3.$ 
 
The inverse limit ${\cal B}_k^{\infty}(V) = \cap {\cal B}_k^r(V)$ of spaces of differential forms is the space of smooth $k$-cochainlets.  These forms are infinitely smooth with a bound on each derivative of order $r$.

\begin{lemma}
${\cal B}_k^{\infty}(V)$ is a Frechet space.
\end{lemma}  

\begin{theorem}
Chainlets form a proper subspace of currents.
\end{theorem}

\begin{proof}
This follows since chainlets form a normed space and currents are just  a topological vector space.
\end{proof} 

Chainlets are a rich and interesting space.  They are more than just a normed space.  They are the direct limit of Banach spaces, and have an orthonormal basis.  So they are ``almost a Banach space''.   They contain a separable, dense inner product space, and are thus ``almost an inner product space''.

    \subsection{Isomorphisms of forms and cochains} By definition,   ${\cal B}_k^r = ({\cal N}_k^r)^*$.   Therefore forms and cochains are canonically isomorphic in this category.
The isomorphism respects the pullback operator since $d f^* = f^* d.$ (See \cite{iso} for a much longer proof of the same result.)  

\begin{theorem}  The spaces ${\cal B}_k^r$ and $({\cal N}_k^r)^*$ are naturally isomorphic.  
\end{theorem}

The de Rham isomorphism of cohomology will later be found as a corollary since singular chains are dense in chainlets in a manifold.  
 
  \subsection{The part of a chainlet in an open set}  Let $U \subset \R^n$ be open.  For a monopolar chain $P = \sum (p_i; \a_i)$, define $P\lfloor_U = \sum  (p_{ij}, \a_{ij})$ where $p_{ij}$ are points in the support of $P \cap U.$  (See \cite{ravello} for more details of this section, including a discussion of nonexceptional open sets.)  Suppose $J$ has finite mass. If $U$ is nonexceptional w.r.t $J$ then if $P_i \to J$ with $M(P_i) \to M(J)$ then  $P_i\lfloor U$ is a Cauchy sequence. We denote the limit by $J\lfloor U.$

 \subsection{Improper chainlets}    
 A sequence of monopolar chains $P_i$  is said to {\itshape \bfseries converge to an  improper chainlet} $J$ if $P_i\lfloor_{U} $ converges to a chainlet, denoted $J\lfloor_{U}$, for each open set $U$ in $V$.
An improper chainlet $K$ is the boundary of an improper chainlet $J$, if $\p (J\lfloor_{U}) - J \lfloor_{\p U}  = K\lfloor_{U}$ for all nonexceptional $U$.  We write $\p J = K$.   An example of an improper chainlet is $\R^n$.  Consider a binary mesh of $\R^n$ with vertices $p_i$.   Let $Q$ denote the unit $n$-cube and define $P_i = \sum 2^{-nk}(p_i; Q).$   It is left to the reader to verify that $P_i$ converges to the improper chainlet $\R^n.$

 \subsection{Operators on chainlets}  Operators on monopolar chains are continuous in the chainlet norms, and therefore extend to operators on chainlets.   
\subsubsection{Prederivative operator $\nabla_u$}  It $J \in {\cal N}_k^r$ is a chainlet, choose $P_i \to J$ in ${\cal N}_k^r$.   	 According to Theorem \ref{contnab}, we know $\nabla_u P_i$ is a Cauchy sequence.  Define  $\nabla_u J = \lim \nabla_u P_i.$  Therefore, $\nabla_u: {\cal N}_k^{r-1} \to {\cal N}_k^{r}$ satisfies  $$|\nabla_u J|^{\natural_{r}} \le |u||J|^{\natural_{r-1}}.$$

The other operators and relations described in Section 3 for monopolar chains extend in a similar fashion, including boundary, pushforward, prederivative and perp.

\subsubsection{Boundary $\p J$}	Define $\p J = \lim \p P_i$.   Since the boundary operator is continuous, it follows that $\p \circ \p =0$ on chainlets, giving us a chainlet complex
$${\cal N}_{k+r}^{0}(V) \buildrel \p \over \to  \cdots \buildrel \p \over \to {\cal N}_{k+1}^{r-1}(V) \buildrel \p \over \to  {\cal N}_k^r(V) \buildrel \p \over \to {\cal N}_{k-1}^{r+1}(V) \buildrel \p \over \to \cdots \buildrel \p \over \to {\cal N}_{0}^{k+r}(V).$$

 \subsubsection{Integration over chainlet domains}  Define  the integral over chainlets $J \in {\cal N}_k^r(V)$ and for  forms $\o$ is of class ${\cal B}_k^r$ as follows:  $$\int_J \o = \lim_{i \to \i}  \o(P_i)$$ where $P_i \to J$ in the chainlet norm.  The limit exists for $J \in {\cal N}_k^{r}$ and $\o \in {\cal N}_k^{r}, 0 \le r \le \i$, since $$|\o(P)| \le |\o|^{\natural_{r}}|P|^{\natural_{r}}.$$

 The support of a chainlet is defined by its complement, using the chainlet integral.  A point $p$ is in the complement of $supp(J)$ if there exists a neighborhood $U$ missing $K$ such that $\int_J \o = 0$ for all forms supported in $U$.

By taking limits, we deduce the following results.
\begin{theorem}\label{oldintegral} If $J \in {\cal N}_k^r, 0 \le r \le \i,$ and $\o \in {\cal B}_k^r$, then
$$ \left|\int_{J} \o \right| \le  |J|^{\natural_{r}}|\o|^{\natural_{r}}.$$
\end{theorem}

The next  result   is a primitive form of Stokes' theorem as it relates   prederivatives of chainlets to directional derivatives of forms. 
  
\begin{theorem}\label{pred}  For $\o \in {\cal B}_k^{r+1}$ and $J \in {\cal N}_k^r,$ $0 \le r \le \i$,
$$\int_J D_u \o = \int_{\nabla_u J} \o.$$
\end{theorem}

    The following general form of Stokes' theorem extends the classical version and applies to all chainlets $J$, including soap films, polypolar chains, and fractals.  
\begin{theorem}[Chainlet Stokes' theorem]\label{SJ} For $\o \in {\cal B}_k^{r+1}$ and $J \in {\cal N}_{k+1}^r,$ $0 \le r \le \i$,

$$\int_{\p J} \o = \int_J d \o.$$
\end{theorem}

\begin{proof}
$$\int_J d \o = \lim \int_{P_i} d\o =  \lim \int_{\p P_i} \o = \int_{\p J} \o.$$
\end{proof}

 \begin{theorem}[Chainlet change of variables] \label{change} For $\o \in {\cal B}_k^{r},$ $J \in {\cal N}_{k}^r,$ and $f \in {\cal B}^r$, $0 \le r \le \i$,
$$\int_{J} f^* \o = \int_{f_*J} \o.$$

\end{theorem}

\begin{proof}
$$\int_{J} f^* \o  = \lim \int_{P_i} f^* \o = \lim \int_{f_* P_i} \o = \int_{f_* J} \o.$$
\end{proof}
 
\begin{theorem}[Chainlet star theorem]
$$\int_J \star \o = \int_{\perp J} \o.$$
\end{theorem}

\begin{corollary}[Chainlet divergence theorem]\label{div}
$$\int_{ J}  d \star \o = \int_{\perp \p J} \o.$$
\end{corollary}

\begin{corollary}[Chainlet curl theorem]
$$\int_J \star d \o = \int_{\p \perp J} \o.$$
\end{corollary}
  
\subsection{Four dense subsets of chainlets}
 \subsubsection{Monopolar chains}   Monopolar chains are dense in chainlets, of course.    Monopolar chains ${\cal P}$ are not only  a Banach algebra but they are also an inner product space,  both  structures not present in chainlets.  But the inner product gives us certain numerical advantages for chainlets, since an orthonormal basis for ${\cal P}$ is also an  orthonormal basis of each ${\cal N}_k^{r}, 0 \le r \le \i.$    \paragraph{\itshape \bfseries Inner product on monopolar $k$-chains}Define $$<P, Q> =  \int_{P \wedge \star Q} dV.$$  The norm $|P|_2 =  \sqrt{<P,P>} =  \sqrt{\sum  M(\a_i)^2}$    is an ``$L^2$ norm'' for monopolar chains.   For an ``$L^p$ norm'',  $p \ge 1,$ define $$|P|_p = (\sum  M(\a_i)^p)^{1/p}.$$

One may extend the  geometric product of Hestenes simply by taking the sum $PQ = <P,Q> + P \wedge Q$, $P, Q \in {\cal P}_k^0.$  Then $PQ$ is a well-defined element of the algebra of monopolar chains ${\cal P}.$   It is an open question whether this will lead to anything more than chainlets already present.   A forthcoming extension of the theory to Clifford algebras, however, is clearly deeply important.

  \subsubsection{Polyhedral chains} Polyhedral chains were invented by Whitney and were the basis to his Geometric Integration Theory \cite{whitney}.  We show that polyhedral chains are naturally included in the space of chainlets as a dense subset.     
  \begin{proposition} 
A $k$-cell $\s$ naturally corresponds to a chainlet.  
\end{proposition}

\begin{proof} First consider a $k$-cube $Q$.   Divide it into subcubes $Q_i$.  Let $p_i$ be the midpoint of $Q_i.$   Let $(p_i; \a_i)$ be the unique   $(0,k)$-pole supported in $p_i$ with mass the same as the mass of $Q_i$ and the same $k$-direction as that of $Q$.  The monopolar chain $\sum (p_i; \a_i)$ is Cauchy in the natural norm. (See \cite{ravello}.)  Call the limit $J_Q$.  Finally, if $\s$ is a cell, subdivide it into its Whitney decomposition of cubes.   The sum converges in the mass norm to a chainlet $J_{\s}.$ 
\end{proof} 
Later, we will see that   $\int_{J_\s} \o = \int_{\s} \o$ in the classical setting, where the LHS is the chainlet integral of $\S \ref{chainletcomplex}$ and the RHS is the Riemann integral.

A polyhedral chain is defined to be a chain of $k$-cells.  
$P = \sum a_i \s_i.$  Every polyhedral chain is therefore a chainlet.

\begin{theorem}
Polyhedral chains are dense in chainlets.
\end{theorem}

\begin{proof}
This follows since monopolar chains are dense in chainlets and polyhedral chains are limits of monopolar chains.
\end{proof}

Polyhedral chains include simplicial chains which are important in algebraic topology.  Singular chains are also dense in chainlets. There is a natural chainlet Mayer-Vietoris result for chainlet homology using operators in this paper which  will give a method for computing chainlet homology classes.   
  \subsubsection{Smooth submanifolds  with boundary in $\R^n$}
  An oriented, smooth $k$-submanifold $N$ in $\R^n$ may be triangulated.    Each simplex $\s_i$ of the triangulation determines a $k$-direction and a $k$-mass. Let $\a_i$ be the unique $k$-cell with the same mass and $k$-direction.   Choose a point $p_i$ in the simplex, or near it.  The monopolar chain  $\sum (p_i; \a_i)$ is Cauchy in the $1$-natural norm.  The limit chainlet $J_N$ represents $N$ in that $\int_N \o = \int_{J_N} \o$ for all forms $\o \in {\cal B}_k^1.$  (The LHS is the Riemann integral over submanifolds, and the RHS is the chainlet integral.) Smooth submanifolds are dense in chainlets since polyhedral chains are dense.

 \subsubsection{Differential forms}  We embed ${\cal B}_k^r$ in ${\cal N}_k^r$.   
 A $k$-form $\o$ is defined as a linear functional of monopolar chains.   By the  Riesz representation theorem, for each $p \in V$, $\o$ at $p$ is represented by a $k$-vector $\b$ such that $\o(p; \a) = <\b ,\a>$  for all simple $k$-vectors $\a.$  Recall $P_i = \sum 2^{-nk}(p_i; Q)$ converges to the improper chainlet $V$.  Define $B_i = \sum 2^{-nk}(p_i; \b_i)$ where $\b_i = \b(p_i).$  Then $B_i$ converges to an improper chainlet $J_{\o}.$  Furthermore,

\begin{lemma} $$\int_{J_{\o}} \eta = \int \eta \wedge \star \o$$ and $|J_{\o}|^{\natural_{r}} = |\o|^{\natural_{r}}.$   
\end{lemma}

\begin{proof}
$$\int_{J_{\o}} = \lim \int_{B_i} \eta  = \lim \sum 2^{-nk} \eta(p_i; \b_i) = \lim \sum 2^{-nk} \eta \wedge \star \o(p_i)  = \int \eta \wedge \star \o.$$ 
\end{proof}
 
 \section{Measure theory and chainlets}
\subsection{Lower semicontinuity}  Define $$|\o|^{\natural_{r}}_{\rho} =
\max\{\|\o\|_0, \dots, \rho\|\o\|\}.$$  Then the norms $|\o|^{\natural_{r}}_{\rho}$ are decreasing as $\rho \to 0$ to the norm  $|\o|^{\natural_{r-1}}.$ It follows that the dual norms $|P|^{\natural_{r}}_{\rho}$ are increasing to the norm $|P|^{\natural_{r-1}}$.   The Banach spaces $|P|^{\natural_{r}}_{\rho}$ are isomorphic to ${\cal N}_k^r$ for each $\rho.$ (The spaces are the same, only the norms are different.) So each norm $|P|^{\natural_{r}}_{\rho}$ is continuous in ${\cal N}_k^r$.  This proves that the limiting norm, the $(r-1)$-natural norm, is lower semi-continuous in ${\cal N}_k^r$.   We similarly show that each $|P|^{\natural_{s}} $ is lower semicontinuous in ${\cal N}_k^r$ for each $0 \le s \le r$, by multiplying the higher order terms by $\rho.$

If $J \in {\cal N}_k^r$ and $0 \le s \le r$ define 
$$|J|_s = \inf\{\liminf |P|^{\natural_{s}}: P_i \buildrel \natural_r   \over  \to J\}.$$

This quantity is the $s$-norm in the $r$-natural space.  It coincides with the $s$-natural norm if $J$ is a monopolar chain $P$.  This follows from lower semicontinuity of $|\cdot|_s.$    In particular, we may now freely speak of the {\itshape \bfseries mass} of a chainlet $J$, realizing that the mass might be infinite.  

\begin{theorem}\label{translate}  If $J \in {\cal N}_k^r$ then 
$|T_u J - J|^{\natural_{r}} \le |u||J|_{r-1}.$
\end{theorem}

\begin{proof}
Choose $P_i \to J$ with $|P|^{\natural_{r-1}} \to  |J|_{r-1}.$  The result follows since $|T_u P - P|^{\natural_{r}} \le |u||P|_{r-1}.$
\end{proof}

\subsection{The operator {\itshape \bfseries $Vec^j$ }}  We define an operator $Vec^j:{\cal N}_k^r \to \L_k^j(V)$ as follows.   It will be fundamental to integration, differentiation and measure of chainlet domains.

For $P = \sum (p_i; \a_i^j)$, define $$Vec^j(P) = \sum   \a_i^j.$$

In what follows, set $j = 0$.  The general case (proved in a sequel) is similar, but is not needed for this draft.

\begin{lemma}\label{Riemann}
Suppose $P$ is a monopolar chain   and $\|\o\|_0 < \i.$   If $\o(p) = \o_0$ for a fixed coelement $\o_0$ and for all $p,$ then $$\int_P \o = \o_0 (Vec^0(P)).$$
\end{lemma} 

\begin{proof}
This follows for monopolar chains $P \in {\cal P}_k^0$ since $\o_0(\a) = \o_0(Vec^0(\a)).$   
  
\end{proof}

 \begin{theorem}\label{massr}
If $P$ is a monopolar $k$-chain and  $r \ge 1$, then  $$M(Vec^0(P)) \le |P|^{\nat_r}. $$
If $supp(P) \subset B_{\e}(p)$  for some $p \in V$ and  $ \e > 0$, then 
$$|P|^{\nat_1} \le M(Vec^0(P)) + \e M(P).$$
\end{theorem}  

\begin{proof} Let $\eta_0$ be a coelement such that $|\eta_0|_0 = 1$, and $\eta_0 (Vec^0(P)) = M(Vec^0(P))$. Define the $k$-form $\eta$ by $\eta(\alpha_p)  := \eta_0(\alpha) .$  Since $\eta$ is constant it follows that $\|\eta\|_r = 0$ for all $r > 0$  and $\|d\eta\|_r = 0$ for all $r \ge 0.$  Hence $|\eta|_r = |\eta|_0  = |\eta_0|_0= 1.$
  By Lemma \ref{Riemann} and Theorem \ref{oldintegral} it follows that 
$$M(Vec^0(P)) = \eta_0 (Vec^0(P) ) = \int_P \eta \le    |\eta|_r |P|^{\natural_{r}} = |P|^{\natural_{r}}.$$

For the second inequality we use the definition of the $r$-natural norm.  It suffices to show that $\frac{|\int_P \o|}{|\o|^{\natural_{1}}}$    is less than or equal the right hand side for any $1$-form $\omega$ of class $B^1$.   Given such $\omega$ define the $k$-form        $\omega_0(\alpha_q)  := \o(\alpha_p) $ for all $q$. By Lemma \ref{Riemann} 
$$
\begin{aligned}
 \left|\int_P \o\right| &\le \left|\int_P \omega_0\right| + \left|\int_P \omega -\omega_0\right |  \\&\le
 |\o(p)(Vec^0(P))| + \sup_{q\in supp(P)}|\o(p) -\o(q)| M(P) \\& \le
   \|\o\|_0M( Vec^0(P))  +  \e \|\o\|_1 M(P) \\& \le
   |\o|^{\natural_{1}}( M(Vec^0(P)) + \e M(P))    \end{aligned}
$$
\end{proof}

 \begin{corollary}\label{massvec}
If $J$ is a   $k$-chainlet of class ${\cal N}^r$,  $r \ge 1,$ then  $$M(Vec^0(J)) \le |J|^{\nat_r}. $$
 \end{corollary}
\begin{proof}
By Theorem \ref{massr} the result holds for monopolar chains.  Choose $P_i \to J$ with $M(P_i) \to M(J).$  Then $Vec^0(P_i) \to Vec^0(J)$ by continuity of $Vec$.  By lower semicontinuity of mass, $$M(Vec^0(J)) \le \liminf M(Vec^0(P_i)) \le  \liminf |P_i|^{\natural_{r}} = |J|^{\natural_{r}}.$$

\end{proof}
 
\begin{theorem}\label{point}
Let $J_i$ be a sequence of $k$-chainlets of class ${\cal N}^r$ such that for some $C > 0$  and $k$-vector $\a$   the following hold: $$M(J_i) < C, \, supp(J_i) \subset B_{\e_i}(p),  \, Vec^0(J_i) \to \a \mbox{ as }  \e_i \to 0.$$  Then $J = \lim J_i$ exists and $Vec^0(J) = \a.$
\end{theorem}

\begin{proof}  Suppose $\e_j \le \e_i$ for $j > i$.   Choose monopolar chains $P_s \to J_i -J_j$ with $M(P_s) \to M(J_i - J_j)$  and $supp(P_s) \subset B_{\e_j}(p)$.   By Theorem \ref{massr} 
 $$|P_s|^{\nat_1} \le M(Vec^0(P_s)) + \e M(P_s) .$$  Hence
$$|J_i - J_j|^{\natural_{r}} \le |J_i - J_j|^{\natural_{1}} \le 
 \liminf M(Vec^0(P_s)) + \e_i M(J_i -J_j).$$
 But $Vec^0(P_s) \to Vec^0(J_i - J_j) \to 0$ by continuity of $Vec$.   
Hence $J = \lim J_s$ exists and $Vec^0(J) = \a.$
 \end{proof}

\begin{corollary}
$Vec^0(\p J) = 0.$
\end{corollary}

\begin{proof}
It suffices to prove this for $(0,k)$-poles $\b$, by continuity of the operators.  But $Vec^0(p;\nabla_u \a) = 0$ implies $Vec^0(\p \b) = 0.$
\end{proof}

Given a monopolar chain $P = \sum a_i \a_i$, let $\l P$ denote its normalized linear contraction:  $ \l P = \sum   (\l p_i; \l^k \a_i)/\l^k.$  
This contraction operator is continuous and extends to chainlets.    By Theorem \ref{point} $Vec^0(J) = \lim_{\l \to 0} \l J.$

\subsection{Part of a chainlet in a Borel set}
 The norm on a finitely additive $k$-vector valued set function $\mu$ is defined as   $$|\mu|^{\natural_{r}} = \sup \left\{\int_{V} \o \cdot d\mu: |\o|^{\natural_{r}} \le 1\right\}.$$
 
 We can extend Whitney's theorem  (\cite{whitney}, XI Theorem 11A) to associate a $k$-chainlet $J$ with finite mass to a finitely additive $k$-vector valued set function $\mu_J$ defined by $$\mu_J(X) = Vec^0(J\lfloor_X).$$    

\begin{theorem}[Chainlet representation theorem]\label{measure}  If $J$ is a $k$-chainlet with finite mass, there exists a unique $k$-vector valued set function $\mu_J$ such that 
$$\int_J \o = \int_V \o \cdot d\mu.$$   The correspondence is an isomorphism such that $$|\mu_J|^{\natural_{r}} = |J|^{\natural_{r}}.$$
\end{theorem}

Whitney's proof extends, but the methods in this paper lead to a more direct proof.  

The ideas extend to define $j$-polar $k$-vector valued set functions, leading to a similar representation theorem for chainlets with $|J|_j < \infty.$ 
 
  The isomorphism preserves operators.  
\subsection{Numerical method} Fix a mesh of $V$.   Let $J$ be a chainlet with finite mass in $V$.  Let $U_i$ be a cell in the mesh.  We can define the part of $J$ in $U_i$ and denote it by $J|_{U_i}$.  Then $Vec^0(J|_{U_i})$ is a $k$-vector. Let $p_i$ be a point in $U_i.$ The sum $\sum  Vec^0(J|_{U_i})$ converges to $J$ in the natural norm.  That means we may do calculus over this monopolar chain and be assured that our results converge to the continuum limit.
 The operators involved in defining these monopolar chains all may be implemented numerically.

\subsection{Graphs of $L^1$ functions} (sketch) A nonnegative  $L^1$ function $f:[0,1] \to \R$ is an increasing limit of   simple functions  $g_i$ each of which is a sum of indicator functions satisfying $$\int \xi_S d\mu = \mu(S)$$ where $\mu$ denotes Lebesgue measure. The graph of an indicator function $\xi_S$ is a $1$-chainlet with finite mass by Theorem \ref{measure}.    Denote the graph of a simple function $g$ by $\G$.  The chainlet  $\G_i - \G_k$ is a finite sum of difference chainlets.  We estimate the total $1$-norm of these difference chainlets.       Since the Lebesgue area of the subgraph of $f$ is finite, the $1$-norm of $\G_i - \G_k$ tends to zero as $i, k \to \i.$ (Use Theorem \ref{translate}.)

  	\section{Vector fields and chainlets}  

  \subsection{Multiplication by a function}   A zero form $\phi \in B_0^r$ is a function $\phi: V \to \R.$   We may write $\phi(p) = \phi(p; 1),$ depending on the context.   Recall  $|\phi|^{\natural_{r}} = \max\{\|\phi\|_0, \dots, \|\phi\|_r\}.$

Define $\phi \sum (p_i; \a_i) =  \sum (p_i; \phi (p_i) \a_i).$

\begin{proposition}\label{longlist}
\begin{enumerate}
\item $\phi(P + Q) = \phi(P) + \phi(Q)$
\item $(\phi + \psi)P = \phi P + \psi P$
\item $(\phi \psi)P = \phi(\psi P)$
\item $\phi P = a P$ if $\phi(p) = a$ for all $p$
\item $M((\phi + \psi) P) = M(\phi P) + M(\psi P)$ if $\phi(p), \psi(p) \ge 0$
\item $M(\phi P) \le M(\psi P)$ if $0 \le \phi(p) \le \psi(p).$ 

\end{enumerate}
\end{proposition}

\begin{proof}
These follow readily from the definitions.  \end{proof}

\begin{proposition}\label{whit}
\begin{enumerate}
\item $M(\phi \p P - \p \phi P) \le k|\phi|^{\natural_{1}}M(P)$
\item $M( \p \phi P) \le k |\phi|^{\natural_{1}} M(P) + |\phi|^{\natural_{0}} M(\p P).$
\end{enumerate}
\end{proposition}

\begin{proof}
These are essentially the same as in \cite{whitney}, p. 209.
\end{proof}

\begin{theorem}\label{hard}  If $\phi \in {\cal B}_0^r$ and $P \in {\cal P}_k$, then 
$$|\phi P|^{\natural_{r}} \le \sum_{i =0}^r \left(\begin{array}{c}r \\i\end{array}\right)|\phi|^{\natural_{i}} |P|^{\natural_{r}},$$
\end{theorem}

\begin{proof}
Let $\e > 0.$  By Theorem \ref{method}, there exists $P = \sum_{j=0}^r D^j$  such that $D^j \in {\cal P}_k^j$ and $|P|^{\natural_{r}} > \sum \|D^j\|_j -\e.$    It suffices to prove that $$|\phi D^j|^{\natural_{j}} \le   \sum_{i=0}^j \left(\begin{array}{c}j \\i\end{array}\right) |\phi|^{\natural_{i}} \|D^j\|_j.$$  Clearly, this holds for $j = 0$.   
Assume it holds for norms less than $j$.  Let $U = (u, U')$ and $\b = \Delta_{U'}^{j-1} \a.$  Then $$\Delta_U^j \a = \Delta_u(\Delta_{U'}^{j-1} \a) = \Delta_u \b.$$  Therefore,
$$\begin{aligned} |\phi (p; \Delta_U^j \a)|^{\natural_{j}} &= |\phi(p; \Delta_u \b)|^{\natural_{j}} \\&\le |\phi T_u(p; \b) - T_u \phi (p;  \b)|^{\natural_{j-1}} + |T_u \phi(p; \b) - 
\phi(p; \b)|^{\natural_{j}}.
\end{aligned}$$ 
Now 
$$\begin{aligned}|\phi T_u(p; \b) - T_u \phi(p; \b)|^{\natural_{j-1}} &= |(\phi(p+u) - \phi(p))(p+u;  \b)|^{\natural_{j-1}} \\&\le  \sum_{i=0}^{j-1} \left(\begin{array}{c}j-1 \\i\end{array}\right) | T_u \phi-\phi|^{\natural_{i}} |\b|^{\natural_{j-1}} \\&\le  \sum_{i=0}^{j-1}  \left(\begin{array}{c}j-1 \\i\end{array}\right)|\phi|^{\natural_{i+1}} |u| \|\b\|_{j-1} \\&\le\sum_{i=0}^{j-1}  \left(\begin{array}{c}j-1 \\i\end{array}\right)|\phi|^{\natural_{i+1}}\|\Delta_U^j \a\|_j,\end{aligned}$$ and 
$$\begin{aligned} |T_u \phi \b - 
\phi \b|^{\natural_{j}} \le  |u||\phi \b|^{\natural_{j-1}} &\le \sum_{i=0}^{j-1}
 \left(\begin{array}{c}j-1 \\i\end{array}\right) |\phi|^{\natural_{i}} |u||\b|^{\natural_{j-1}}   \\&\le \sum_{i=0}^{j-1} \left(\begin{array}{c}j-1 \\i\end{array}\right) |\phi|^{\natural_{i}} \|\Delta_U^j \a\|_j \end{aligned} .$$
 It follows that
 $$ |\phi (p; \Delta_U^j \a)|^{\natural_{j}} \le \sum_{i=0}^j \left(\begin{array}{c}j \\i\end{array}\right) |\phi|^{\natural_{i}}\|\Delta_U^j \a\|_j.$$
\end{proof}

  Define $\phi \o(p; \a) = \o(p; \phi(p) \a).$ 
 It follows by linearity that  $$ \phi \o(P) = \o(\phi P).$$

The next result follows from Proposition \ref{longlist}.

\begin{proposition}
\begin{enumerate}
\item $\phi(\o + \eta) = \phi(\o) + \phi(\eta)$
\item $(\phi + \psi)\o = \phi \o + \psi \o$
\item $(\phi \psi)\o = \phi(\psi \o)$
\item $\phi \o = a \o$ if $\phi(p) = a$ for all $p$
\item $|(\phi + \psi) \o|^{\natural_{0}} = |\phi \o|^{\natural_{0}} + |\psi \o|^{\natural_{0}}$ if $\phi(p), \psi(p) \ge 0$
\item $|\phi \o|^{\natural_{0}} \le |\psi \o|^{\natural_{0}}$ if $0 \le \phi(p) \le \psi(p)$ 
 
\end{enumerate}
\end{proposition}

  The next two propositions follow from Proposition \ref{whit} and Theorem \ref{hard}.

\begin{proposition}
\begin{enumerate}
\item $|\phi d \o - d \phi \o|^{\natural_{0}} \le (k+1)|\phi|^{\natural_{1}}|\o|^{\natural_{0}}$
\item $| d \phi \o|^{\natural_{0}} \le (k+1) |\phi|^{\natural_{1}} |\o|^{\natural_{0}} + |\phi|^{\natural_{0}} |d \o|^{\natural_{0}}.$
\end{enumerate}
\end{proposition}

\begin{proposition} 
$$|\phi \o|^{\natural_{r}} \le \sum_{i =0}^r \left(\begin{array}{c}r \\i\end{array}\right)|\phi|^{\natural_{i}} |\o|^{\natural_{r}}.$$
\end{proposition} 

\subsection{$k$-vector fields}   A $k$-vector field $X$ on $V$ is defined to be a function $X: V \to \L_k(V)$.  Therefore,  functions are a field of monopolar $0$-vectors.   We next consider monopolar $k$-vector fields.

The vector field $e_1$ in $V$ is associated to the $1$-chainlet $J_{e_1}$ defined as follows: We first define  $J_{e_1}$ in a unit $n$-cube $Q$.  Let $Q = R \times I$ where $R$ is an $(n-1)$-face of $Q$ and $I$ is a $1$-cell with direction $e_1.$ Let $p_0$ be the midpoint of $R$.  Subdivide $R$ into binary $(n-1)$-cubes $R_{k,i}$ with side length $2^{-k}$.  Denote the midpoints of the $R_{k,i}$ by $p_{k,i}.$  Let $P_0 = \s_0$ denote the    $1$-cell supported in $Q$, with direction $e_1$, and with one endpoint $p_{0}$.  Let $P_k = \sum 2^{-nk}\s_{k,i}$ with each $\s_{k,i} $ parallel to $\s_0$ and with endpoint $p_{k,i}.$ Then $M(P_k) = 1$ and the sequence $\{P_k\}$ is Cauchy in the $1$-norm since the distance of translation tends to zero and the total mass that is translated i $M(Q).$ (See  Theorem \ref{method} or \cite{gmt} for details.)     Denote the limit by $J_{e_1}$.  Use the preferred basis to construct $n$-chainlets $J_{e_i}$ associated to the vector field $e_i$.   

A vector field $X$ can be written $X = \sum_{i = 1}^n \phi_i e_i$.  Therefore, the chainlet
$J_X = \sum_{i = 1}^n \phi J_{e_i}$ is associated to $X.$   Define $$|X|^{\natural_{r}} = \sum_{i = 1}^n |\phi_i|^{\natural_{r}}.$$

\subsection{Exterior product  $E_X J$} 
 Let $\b \in \L_m(V)$.
Define the exterior product operator $$E_{\b}: {\cal P}_k^j \to {\cal P}_{k+m}^j$$ by $$E_{\b} \sum (p_i; \a_i) = \sum  (p_i;\a_i \wedge \b).$$

The next two results follow directly from the definitions.  
\begin{lemma}\label{lots} $E_{\b} P$ is a bilinear operator in $v$ and $P$ satifsying
\begin{enumerate}
\item $E_{\b} E_{\b} = 0$
\item $E_{\b}(P \cdot Q) = E_{\b} P \cdot Q + (-1)^k P \cdot E_{\b} Q$
\item $f_* E_{\b} P = E_{f_*\b} f_* P$
\item $\phi E_{\b} P = E_{\b} \phi P = E_{\phi \b} P$
\item $|E_{\b} P|^{\natural_{r}} = |P|^{\natural_{r}}.$
\end{enumerate}
\end{lemma}
\begin{proposition}
$$|E_{\b} P|^{\natural_{r}} \le M(\b)|P|^{\natural_{r}}.$$
\end{proposition} 

If $f: M \to N$ is a mapping and $Y$ is $f$-related to $X$, that is, $f_* \circ X = Y \circ f$, then  
$E_Y( f_* P) = f_* (E_X P).$ Of course, is $f$ is a diffeomorphism, then $f_* X$ is always $f$-related to $X$.

Define the {\itshape \bfseries interior product} operator of forms by duality $$i_{\b} : {\cal B}_{k+m}^r \to {\cal B}_k^r$$ as  $$i_{\b} \o(p;\a) = \o(p; E_{\b} \a).$$ Lemma \ref{lots} yields
 \begin{lemma} $i_{\b} \o$ is a bilinear operator in $v$ and $\o$ satisfying
\begin{enumerate}
\item   $i_{\b} i_{\b} = 0$
\item $i_{\b}(\o \wedge \eta) = i_{\b} \o \wedge \eta + (-1)^k \o \wedge i_{\b} \eta$
\item $f^* (i_{\b} \o) = i_{f_*\b} f^* \o$
\item $\phi i_{\b} \o = i_{\b} \phi \o = i_{\phi \b} \o;l$
\item $|i_{\b} \o|^{\natural_{r}} = |\o|^{\natural_{r}}.$
\end{enumerate}
\end{lemma}

\begin{theorem} Let $X$ be an $m$-vector field on $V$, and $P$ a monopolar $k$-chain.  There exist constants $C(X,k,m, r) > 0$ such that 
$$|E_X(P)|^{\natural_{r}} \le C(X,k,m, r)|X|^{\natural_{r}}
|P|^{\natural_{r}}.$$
\end{theorem}

\begin{proof} Set $X = \sum \phi_i e_i.$  Observe that $|E_{e_i}(P)|^{\natural_{r}} \le |P|^{\natural_{r}}.$  Then

$$ \frac{\o(E_{e_i}P)}{|\o|^{\natural_{r}}}  = \frac{  i_{e_i} \o(P)}{|\o|^{\natural_{r}}} = \frac{i_{e_i} \o (P)}{|i_{e_i} \o|^{\natural_{r}}} \le \sup\frac{\o(P)}{|\o|^{\natural_{r}}} = |P|^{\natural_{r}}. $$

Therefore, by Lemma \ref{lots} (iv) and Theorem \ref{hard}
$$|E_{\phi e_i}(P) |^{\natural_{r}} = |\phi(E_{e_i}P)|^{\natural_{r}} \le C_i(\phi, r)|\phi|^{\natural_{r}} |P|^{\natural_{r}}$$  where the constants $C_i(\phi, r) =  \sum_{i =0}^r \left(\begin{array}{c}r \\i\end{array}\right)|\phi|^{\natural_{i}}.$  Now take sums over the basis $(e_1, \dots, e_n)$. To obtain the result, set $C(\phi,r) = \sum C_i(\phi,r).$
 
\end{proof}  
If $J$ is a chainlet in ${\cal N}k^r$, $X$ and $m$-vector field, and $P_i \to J$, 
define $$E_X J = \lim E_X P_i \in {\cal N}_{k+m}^r.$$ 
Lemma \ref{lots} (iii) extends to $J$:
$$f_* E_X J = E_{f_* X} f_* J$$ and $$f^* i_X \o = i_{f_*X} f^* \o.$$
 A similar construction leads to $$E_{X^j} J  $$ for a $j$-polar $m$-vector field $X^j$.    As with most of the operators,  $j = 1, 2, \dots$, we obtain  fields polypolar vectors expressed as   chainlets.    
 
  Since the operators and products commute with pushforward, the definitions and relations extend to smooth manifolds.

If $f: M \to N$ is a smooth mapping and $Y$ is $f$-related to $X$, then $Y = f_*X$ and
$i_Y f^* \o = f^* i_X \o.$

\subsection{Lie derivative ${\cal L}_X J$}   
 
If $X$ is a vector field and $ \a$ is a $k$-vector field  in a manifold,   we recall the classical Lie derivative of $\a$ in the direction $X$.  Let $g_t$ be the flow of $X$.  Fix $x_0$ and define $${\cal L}_X (\a)_{x_0} := lim_{t \to 0} \frac{g_{-t*} \a_{x_t} - \a_{x_0}}{t}.$$     If $X$ is smooth, then the limit exists.   

We similarly define ${\cal L}_X$ on fields of $(j,k)$-poles   The limit exists in the $j$-natural norm as a field of   $(j,k)$-poles.    This leads to a definition of ${\cal L}_X J$ for a chainlet $J$.  

\begin{lemma}
$${\cal L}_X \p  = \p {\cal L}_X.$$
\end{lemma}
\begin{proof}
$${\cal L}_X \p \a = \lim \frac{f_{-t} \p \a_{x_t} - \p \a_{x_0}}{t} = \p \lim\frac{f_{-t}  \a_{x_t} -  \a_{x_0}}{t} = \p {\cal L}_X \a.$$
\end{proof}

Define $${\cal L}_X \o(p;\a) := \o(p; {\cal L}_X \a).$$

${\cal L}_X \o$ is bilinear in $X, \o$ and 
${\cal L}_X \o \wedge \eta = {\cal L}_X \o \wedge \eta + \o \wedge {\cal L}_X \eta.$

\begin{theorem}\label{two}
\begin{enumerate}
\item $Vec^0(\nabla_X) = {\cal L}_X$ 
\item $\p \nabla_X = \nabla_X \p.$
\end{enumerate}

\end{theorem}

\begin{proof}  The first part follows by continuity of the operator $Vec$.
$$d {\cal L}_X \o(p; \a) = {\cal L}_X \o(p; \p \a) = \o(p; {\cal L}_X \p \a) = \o(p; \p {\cal L}_X \a) = {\cal L}_X d\o(p; \a).$$  The second follows from the definitions.
\end{proof}

\begin{theorem}
$\nabla_X  = \p E_X  + E_X \p.$
\end{theorem}

\begin{corollary}[Cartan's magic formula]
${\cal L}_X \o = d i_X \o + i_X d \o.$
\end{corollary}

\begin{proof} By Theorem \ref{two}
${\cal L}_X P = Vec^0(\nabla_X P) = Vec^0( \p E_X P  + E_X \p P).$  It follows that
$d i_X \o(P) + i_X d \o(P) = \o(\p(P \wedge X)) + \o(\p P \wedge X) = 
\o(\nabla_X P) = 
{\cal L}_X \o (P).$
\end{proof}
 
For a diffeomorphism $f$, $$f^* {\cal L}_X \o = {\cal L}_{f_*X} f^* \o.$$
If $f: M \to N$ is a mapping and $Y$ is $f$-related to $X$, then $${\cal L}_Y f^* \o = f^* {\cal L}_X \o.$$

\subsection{Translation $T_X J$}  The translation operator $T_u(p; \a) = (p+u; \a)$ is continuous in the chainlet norm.  Similar techniques as those given above show that this operator extends to $T_X J$ for smooth vector fields $X$ and chainlets $J$.  

\section{The chainlet complex in a manifold}  
Now set $V = \R^n$.
A {\itshape \bfseries singular $k$-chainlet} $g J $  in  a smooth $n$-manifold $M$ of class ${\cal N}^r$  is a continuous mapping $g:\R^n \to M$ and a $k$-chainlet $J$ in $\R^n$.  
Consider the vector space ${\cal N}_k^r(M)$, generated by all singular $k$-chainlets subject to the relation:    $g(p; \a) =  h(q; \b)$ if and only 
$ (h^{-1}\circ g)_*(g(p; \a)) = h (q; \b),$   for all $(p; \a) \in \R^n \times \L_k^0.$  The boundary operator is well defined on chainlets in a manifold by Corollary \ref{push}. We obtain a chain complex of vector spaces since $\p \circ \p = 0.$ 
 $${\cal N}_{k+r}^{0}(M) \buildrel \p \over \to  \cdots \buildrel \p \over \to {\cal N}_{k+1}^{r-1}(M) \buildrel \p \over \to  {\cal N}_k^r(M) \buildrel \p \over \to {\cal N}_{k-1}^{r+1}(M) \buildrel \p \over \to \cdots \buildrel \p \over \to {\cal N}_{0}^{k+r}(M).$$

Forms of class $B_k^r$ and of class $C_k^r$ are well defined on $M$.
  The relation $f^* \o = \o \circ f_*$  leads to a well-defined integral of smooth forms of class $B_k^r$ over chainlets in ${\cal N}_k^r(M)$.   
  $$\int_J \o.$$

The   operators $\nabla_u$ and $\p$ are well-defined on manifolds since they commute with pushforward of diffeomorphisms. 

 We deduce extensions of     Theorem \ref{pred} and Stokes' theorem   \ref{SJ}.

\begin{theorem} $$\int_J D_u \o = \int_{\nabla_u J} \o$$ for $\o \in B_k^r(M)$ and $J \in {\cal N}_k^{r+1}.$
\end{theorem}
 
\begin{theorem}[Stokes' theorem for chainlets in a smooth manifold]
$$\int_J d\o = \int_{\p J} \o$$ for $\o \in B_k^r(M)$ and $J \in {\cal N}_{k-1}^{r+1}.$
\end{theorem}

 In Riemannian manifolds, the translation operator $T_u$ is defined via the covariant derivative.  The operators $\perp$ and Hodge star are well defined, leading to a divergence theorem for chainlets in Riemannian manifolds. 
 
\section{Integration of rough forms over rough domains}
Given a chainlet $J \in {\cal N}_k^r$ and a form $\o \in {\cal B}_k^s, 0 \le s \le r,$ define $$\int_J \o = \sup\{ \liminf \int_{P_i} \o : P_i \buildrel \natural_r \over \to J\}.$$    

Then $$\left|\int_J \o\right| \le |J|_s |\o|^{\natural_{s}},$$   
$$\left|\int_{J}d \o\right| \le \left|\int_{\p J}  \o\right|,$$  
$$\left|\int_{J}d \star \o\right| \le \left|\int_{\perp \p J}  \o\right|,$$ 
  and
$$\left|\int_{ J}D_u \o \right|  \le \left|\int_{\nabla_u J} \o \right|  \le |u||J|_s |\o|^{\natural_{r}}.$$ 

Whitney's example \cite{nonconstant} of a function nonconstant on a connected set of critical points shows this inequality is sharp.  (See also \cite{norton}.)

 \section{Further operators}

\subsection{Slant product $J/X$}
  Suppose $\a \in \L_k^0$ and $\b \ne 0 \in \L_m^0$   with $0 \le m \le k \le n$.      
Define the {\itshape \bfseries slant product}  $${\cal S}_{\b}\a = \a/\b := (-1)^{k(n-k)}\perp(\b \wedge \perp \a)/M(\b)^2 \in \L_{k-m}^0.$$

If $m = 0$, slant product reduces to division of a $k$-vector by a nonzero scalar.   Therefore, if $k = m = 0$, slant product reduces to division of real numbers in $\R^1$.

\begin{lemma} The slant product
${\cal S}_{\b}: \L_k^0 \to \L_{k-m}^0$ is linear and satisfies  
\begin{enumerate}
\item $Vec^0(\b) \subset Vec^0(\a) \implies \b \wedge  (\a/\b)   = \a$
\item  $Vec^0(\b) \perp Vec^0(\a)  \implies (\a \wedge \b)/\b = \a$ and $\a/\b = 0 \in \mathbb{F}$
\item $\a/\a  = 1$.    
\end{enumerate}
\end{lemma}

\begin{proposition}
$|P/\b|^{\natural_{r}} \le |P|^{\natural_{r}}/M(\b).$
\end{proposition}

\begin{proof} Then
$$|P/\b|^{\natural_{r}} = |\b \wedge \perp P|^{\natural_{r}}/M(\b)^2 \le  |P|^{\natural_{r}}/M(\b).$$ 
\end{proof}

Therefore, the slant operator extends to chainlets in $\R^n$.  

$|J/\b||^{\natural_{r}} \le |J|^{\natural_{r}}/M(\b).$

\begin{lemma} Slant product is a linear transformation $${\cal S}_{\b}: {\cal P}_k^j \to {\cal P}_{k-m}^j$$ satisfying
\begin{enumerate}
\item ${\cal S}_{\b} {\cal S}_{\b} = 0$
\item $f_* \circ {\cal S}_{\b}   = {\cal S}_{f_* \b}\circ f_* $
\item $\phi {\cal S}_{\b}  = {\cal S}_{ \b} \phi.$
\end{enumerate}

\end{lemma}
 
Therefore, slant product extends to chainlets in manifolds.

 If $X$ is a nonzero monopolar $k$-vector field,  slant product is defined at each point.  Let us further assume that the mass of each $k$-vectors is bounded below by a constant $K$.
 
\begin{theorem} If $X \in B^r$, there exists a constant $C(r, X)$ such that
$$ | P/X |^{\natural_{r}}  \le  C(r, X) |X|^{\natural_{r}}|P|^{\natural_{r}}.$$
\end{theorem}

\begin{proof}  By Theorem \ref{hard}
$$|\perp ((\phi e_1) \wedge \perp P)|^{\natural_{r}} = |\phi (e_1 \wedge \perp P)|^{\natural_{r}} \le C(r, \phi)|P|^{\natural_{r}}.$$  
Suppose $X = \sum_{i=1}^n \phi_i e_i.$  Let $C(r, X) = \sum C(r, X_i).$  Then
$$ | P/X |^{\natural_{r}}  \le  \sum | \perp(\phi_ie_i(p_s) \wedge \perp P |^{\natural_{r}}/M(X(p_s))^2  \le K^2 C(r,X)|X|^{\natural_{r}} |P|^{\natural_{r}}.$$

\end{proof} 

It follows that $J/X$ is well defined for a nonzero $k$-vector field $X$.  Order $j$ vector fields are treated in a similar fashion.   
 
For each $\b \in \L_m^0$, define the extrusion operator on forms
$$ext_{\b}: {\cal B}_k^r \to {\cal B}_{k+m}^r$$
$$ext_{\b} \o(p;\a) = \o(p;  \a/\b).$$

This extends to $k$-vector fields $X$.  

$$ext_X \o(p; \a) = \o(p; \a/X(p)).$$

Define ${\cal H}_\b = \d ext_{\b} + ext_{\b} \d$ where $\d = \star d \star.$  
It is an open question whether or not   ${\cal H}_\b$ is a derivation. 
 
 Slant product is not generally defined on smooth manifolds since pushforward does not commute with the perp operator.  
 
  \subsection*{Cross product $J \times X$}   
  
  Define $$\times: {\cal P}_{k_1}^{j_1} \times {\cal P}_{k_2}^{j_2}  \to {\cal P}_{n-k_1-k_2}^{j_1+j_2}$$ by
  $$\a \times \b := \perp(\a \wedge \b).$$
  For $k_1 = k_2 = n/3$ this product combines pairs of $(0,k)$-poles   and produces a $(0,k)$-pole.   Of course, when $n= 3$ this corresponds to the standard cross product of vectors. 
   
In the degenerate case with $v = w$, then $Vec^0(v) \wedge Vec^0(w)$ is zero.  Thus  $\perp$ of it is zero.  On the other hand, if $v$ is orthogonal to $w$, then the wedge product is a $(0,2)$-pole with mass the same as the product $|v||w|$, so its perp corresponds to a vector with norm $|v||w|$ that is orthogonal to this $(0,2)$-pole.    

We next show this extends to $J \times X$ where $X$ is a smooth $k$-vector field in $V$.
\begin{theorem}
$$|P \times X|^{\natural_{r}} \le C(r,X)|P|^{\natural_{r}}|X|^{\natural_{r}}.$$
\end{theorem}

\begin{proof} Let $X = \phi_i e_i.$  Then by definition and Theorem \ref{hard},
$$\begin{aligned}|P \times \phi_i e_i|^{\natural_{r}} &= \sum |\perp((p_s, \a_s) \wedge \phi_i(p_s) e_i)|^{\natural_{r}}  \\&= \sum |\phi_i(p_s)(p_s, \a_s \wedge e_i)|^{\natural_{r}} \\&\le C(r, \phi_i)|\phi_i|^{\natural_{r}}|P|^{\natural_{r}}.\end{aligned}$$ The result follows by setting $C(r,X) = \sum C(r, \phi_i).$

\end{proof}

 As usual, we define a dual operator on forms as
 $$(\o \times \b)(p; \a) = \o(p; \a \times \b).$$

\subsection*{Intersection product $J \cap X$}     Let $k_1 + k_2 \ge n.$  Let $\a \in {\cal P}_{k_1}^{j_1}$,  $\b \in {\cal P}_{k_1}^{j_2}$.  Define
$$\a \cap \b := \perp(\perp \a \wedge \perp \b).$$
 
For $j_1 = j_2 = 0$ this identifies the {\itshape \bfseries intersection product} of $\a$ and $\b$.

As above, we may extend this to $J \cap X$ where $X$ is a $k$-vector field.  The dual operator on forms is define by $$\o \cap X(p; \a) = \o(p; \a \cap X).$$

\subsubsection*{Projection}  If $\a \in {\cal P}_{k+j}$, $\b \in {\cal P}_k$ define $$\pi_{\b} \a :=  \perp(\perp \a \cap \b) \cap \b.$$
As above, we may extend this to $\pi_X J$ where $X$ is a $k$-vector field.  The dual operator on forms is define by $$\pi_X \o(p;\a) = \o(p; \pi_X \a).$$ 

\section{Applications}\label{applications}

  \subsection{Solutions to Plateau's Problem}  
  
  Suppose $\s$ is a $2$-cell in $3$-space.  Let  $u$ be a unit vector orthogonal to $Vec^0(\s)$.   Then $\nabla_u \s$ is a dipole surface that locally models a soap film without branches.  By adding three of these along a common dipole edge at angles of $120^{\deg}$, we can obtain a branched surface.  

A mass cell is defined to be $e_u \s$.  Then $\p e_u \s =  \nabla_u \s + e_u(\p \s).$  Sums of mass cells and dipole cells give models of soap films with curvature.  

We call a chain of dipole $k$-cells and mass $k$-cells a $k$-dipolyhedron.   

As an application of chainlet methods, we observe that any naturally arising soap film $S$ spanning a smooth curve $\g$ can be expressed as a limit of dipolyhedra $S = \lim D_i$ in the natural norm with $\p D_i$ supported in $\g$.  The boundary of $S$ is also supported in $\g$.   

A similar construction in $\R^4$ \cite{soap}  leads to existence of minimal spanning set for a fixed Jordan curve,  a solution to Plateau's problem, assuming a bound on energy.  The minimizer has soap film regularity a.e. \cite{plateau} and   has surface area smaller than any other Plateau solutions to date. 

The new methods of this paper shed some light on the general problem which is under investigation.

\subsection{Chainlets and distributions}   The question arises whether or not we may replace currents with the smaller space of chainlets in analysis. The simplest case is $k = 0.$   In this section we ask whether we can ``take the derivative'' of $0$-chainlets.  Distributions are defined over test functions with compact support.   For $0$-chainlets, test functions need only be integrable, in order to take the derivative

If $J$ is a $0$-chainlet in $\R$, define $J' = \nabla_{u}J$  where $u$ is the unit vector $e_1.$   This corresponds to the derivative of a smooth function $f$.
A smooth function $f$ determines a chainlet  $J_f$ if its integral is finite.  \begin{lemma}
$\nabla_u J_f =  -J_{f'}.$
\end{lemma}  

\begin{proof}
The proof reduces to showing $$\int_{\nabla_u J_f} g = -\int_{J_{f'}} g$$ for all  smooth $g$.
That is, $$\int_{J_f} \nabla_u g = \int f g' =- \int f' g.$$   We use integration by parts.  $f(b)g(b) - f(a) g(a) = \int fg' + \int f'g.$  Since the integral of $f$ is finite and $f$ is smooth, the left hand side tends to zero for $a \to \infty$ and $b \to -\infty.$   
\end{proof}  

Chainlets offer more structure, and therefore a richer theory that is easier to work with than   distributions since they form a normed space.     
Applications to PDE's are anticipated.  
 
 \subsection{Dual mesh convergence}  In the discrete theory of simplicial complexes, the problem of existence and convergence of a geometric Hodge star operator has eluded mathematicians until recently.   Barycentric subdivisions or 
or barycentric duals are commonly used.  The circumcentric dual is introduced in \cite{marsden} which has the advantage of orthogonality to the primary mesh.   But convergence to the smooth continuum remained elusive.  The author proved the chainlet discrete Hodge star converged to the smooth continuum (\cite{ravello}, see also \cite{gmt}).  Wilson \cite{wilson} constructed a combinatorial Hodge star operator via a dual mesh,
and showed convergence to the continuum as the mesh
of a triangulation tends to zero.  The methods of \cite{ravello}  apply to simplicial complexes as a special case, but there is no mention there of dual complexes.  We address this now.

 Let $(S, T)$ be an ordered pair of two simplicial complexes.  We will call $(S, T)$ a {\itshape \bfseries   mesh pair} if the $k$-simplexes $\s_k$ of $S$ are in $1-1$ correspondence with the $(n-k)$-simplexes $\t_{n-k}$ of $T$.   We call $S$ the {\itshape \bfseries primary mesh} and $T$ the {\itshape \bfseries dual mesh}.  Let $p_k$ be the barycenter of $\s_k$ and $q_{n-k}$ the barycenter of $\t_{n-k}.$  Let $\a_{n-k} = \perp Vec^0(\s_k)$ and $\b_{n-k} = Vec^0(\t_{n-k}).$    We say that a sequence of mesh pairs $(S^i, T^i)$ is  a {\itshape \bfseries Hodge sequence} if
\begin{enumerate}
\item  The mesh sizes of both  $S^i$ and  $T^i$ tend to zero as $i \to \i.$
\item    $$\frac{M(\b^i_{n-k}- \a^i_{n-k})}{M(\b^i_{n-k})} \to 0, \mbox{ as } i \to \i$$
\item   $$|p^i_k - q^i_{n-k}| \to 0, \mbox{ as }  i   \to \i.$$
\end{enumerate} 

\begin{lemma} If $(S^i, T^i)$ is  a {\itshape \bfseries Hodge sequence} then
$$\frac{|\t_{n-k}^i - \perp \s_k^i |^{\natural_{1}}}{M(\t_{n-k}^i)} \to 0 \mbox{ as } i \to \i.$$
\end{lemma}

\begin{proof}
         
     By the triangle inequality, $$\begin{aligned}|\t_{n-k} - \perp \s_k|^{\natural_{1}}  &\le   |\t_{n-k} - T_{p_k - q_{n-k}} \t_{n-k}|^{\natural_{1}} \\&+  |T_{p_k - q_{n-k}} \t_{n-k}   - (p_k, \b_{n-k})|^{\natural_{1}} \\&+ |(p_k, \b_{n-k})-(p_k, \a_{n-k})|^{\natural_{1}} \\&+ |(p_k, \a_{n-k}) - \perp \s_k|^{\natural_{1}}\end{aligned}.$$  
     We know $$\frac{ |(p_k^i; \a^i_{n-k}) - \perp \s^i_k|^{\natural_{1}}}{M(\s_{k}^i)} \to 0$$  by the definition of the operator $\perp.$  In particular, the distance of translation of estimates of $(p^i_k, \a^i_{n-k}) - \perp \s^i_k$ is tending to zero, the total mass is kept constant.  (See Theorem \ref{method}.)  The second term is similar.  
   
 Also, $$  |\t^i_{n-k} - T_{p^i_k - q^i_{n-k}} \t^i_{n-k}|^{\natural_{1}}  \le |p^i_k - q^i_{n-k}|M(\t^i_{n-k}).$$  
 
 The third term simplifies to 
$$|(p_k^i; \b^i_{n-k}- \a^i_{n-k})|^{\natural_{1}} \le M(\b^i_{n-k}- \a^i_{n-k}).$$

\end{proof}

   Since $\perp$ is a continuous chainlet operator and dual to Hodge star $\star$ on forms, this lemma guarantees that a  Hodge sequence will   limit to the smooth continuum w.r.t. the Hodge star and boundary operators.  Therefore, discrete methods relying on such meshes will produce reliable discrete approximations to Gauss and Stokes' theorems.  Numerical estimates on convergence may be gleaned from the definition and lemma.
 
\subsection{Wavelets over chainlets}  
A {\itshape \bfseries chainlet transform} is defined on $\R^n$.  (Sketch)  Cover $\R^n$ with a binary grid ${\cal Q}$.  Let $J$ be a $k$-chainlet. 
  For each $Q_m$ in the grid,  the part of $J$ in $Q_m$ is a chainlet $J \lfloor_{Q_m}$.   Since mass is lower semi-continuous in the chainlet norm \cite{ravello}, $a_m = M(J\lfloor_{Q_m})$ is well-defined.   Let $p_m$ denote the midpoint of $Q_m$.  Let   ${\cal T}(J, Q) = \sum a_m Vec^0( J\lfloor_{Q_m}).$  Then  ${\cal T}(J, Q)$ is a sequence of monopolar chains converging to $J$ in the $1$-natural norm.   
  
  By setting $P_0 =    {\cal T}(J, Q_0)$ and $P_i = {\cal T}(J, Q_{i+1}) - {\cal T}(J, Q_i),$ we may write  $$J = \sum_{i = 1}^{\i} P_i.$$
  
The $P_i$ may be written in terms of an orthonormal basis of the subspace of monopolar chains $P_i =  \sum_{s= 1}^{\frac{n!}{k!(n-k)!}} B_{ij}$, so that $$J = \sum_{i=1}^{\i} \sum_{s= 1}^{\frac{n!}{k!(n-k)!}} B_{ij}.$$
  
  Wavelets over chainlets now become natural, as do Fourier series over chainlets.  If $\o = \sum_{j = 1}^{\i} \eta_j$ is a wavelet series or Fourier series, then $$\int_J \o = \int_{\sum_{i = 1}^{\i} P_i} \Sigma_{j = 1}^{\i} \eta_j =  \sum_{i = 1}^{\i}\sum_{j = 1}^{\i} \int_{P_i} \eta_j.$$ We end up with a convergent double series of trivial integrals.   All of the operators involved may be implemented numerically since monopolar chains are dense in chainlets.
  \section*{Appendix A}\label{appendixA}

\subsection{Deduction of the coordinate calculus}   We next see that standard integral equations of coordinate calculus follow readily from the concise integral equations (i)-(iii) of chainlets.   This paper, therefore, provides the foundations for a full theory of calculus starting from basic principles of multilinear algebra.\footnote{Indeed, in the spring of 2006, the author will be teaching an experimental course on chainlet theory to 30 math students at Berkeley, all of whom have studied abstract linear algebra, with no assumptions of calculus, algebra, or real analysis.} The following proofs are not possible from the perspective of the Cartan theory which is missing the prederivative operator $\nabla_u$ and the geometric Hodge star operator $\perp$.  We first retrieve the standard formulation of the exterior derivative of a multivariable function $f:\R^n \to \R$:  Let $(e_1, \dots, e_n)$ denote the standard basis of $\R^n$.   The pair $(x; v)$ denotes a $1$-vector $v$ in the tangent space of $x \in \R^n$.       
A {\itshape   partial derivative} of $f$ at $x \in V$ is defined as a directional derivative $$ \frac{\p f}{\p x_i}(x) = D_{e_i}f(x; 1)$$  where $(e_1, \dots, e_n)$ is a preferred basis and $x = x_1 e_1 + \cdots x_n e_n.$

\begin{theorem}
$df(x; e_i)   =  \frac{\p f}{\p x_i}dx_i(x; e_i) .$
\end{theorem}

\begin{proof}
$$\begin{aligned}df(x; e_i) = f (x;  \p e_i) =     f  (x;   \nabla_{e_i}(1)) =   D_{e_i}f (x; 1).\end{aligned}$$  
\end{proof}
  
\begin{corollary}
$df = \sum  \frac{\p f}{\p x_i} dx_i.$
\end{corollary}

The Fundamental Theorem of Calculus follows from (i):   
$$\int_a^b f'(x) dx = \int_{[a,b]} df = \int_{\p [a,b]} f = \int_{\{b\} -\{a\}} f =  f(b) - f(a).$$
Green's theorem over a bounded open set $U \subset \R^2$ also follows from (i):

$$\int_{\p U} Pdx + Q dy =    \int_U  d( P dx + Q dy) =      \int_U \frac{\p Q}{\p x}dxdy - \frac{\p P}{\p y}dx dy.$$

The chainlet divergence theorem follows from (i) and (iii):  $$\int_J d \star \o = \int_{\perp \p J} \o.$$  For a simple domain $V$ in $\R^3$, we retrieve the Gauss' divergence theorem for $1$-forms $F$ defined in a neighborhood of $V$.   Suppose $F = P dx + Q dy + Rdz$.  Then $$\begin{aligned} d \star F &= d\star (Pdx + Q dy + Rdz) =  d(P dydz + Q dzdx + R dxdy) \\&= \left(\frac{\p P}{\p x} +  \frac{\p Q}{\p y} + \frac{\p R}{\p z}\right) dxdydz = \nabla \cdot F dV.\end{aligned}$$  Thus
$$ \int_V \nabla \cdot F dV = \int_V d \star F  = \int_{\perp \p V} F  = \int_{\p V} F \cdot \vec{n}dS.$$
The chainlet curl theorem takes the form $\int_J \star d \o = \int_{\p \perp J} \o$.  Therefore, if $S$ is a surface in $\R^3$ and   $F$ a smooth $1$-form defined in a neighborhood of $S$ in $\R^3$ we calculate 
$$\begin{aligned} \star d F &= \star d(Pdx + Q dy + Rdz)  \\&=  
 \left(\frac{\p R}{\p y} - \frac{\p Q}{\p z}\right)dx + \left(\frac{\p P}{\p z}- \frac{\p R}{\p x}\right)dy 
+   \left( \frac{\p Q}{\p x} -\frac{\p P}{\p y}\right) dz \\&= \nabla \times F \end{aligned}.$$  Therefore,  
 
$$\int_S (\nabla \times F) \cdot \vec{n} dA = \int_{\perp S} \star d F =  \int_{\p S} F =  \int_{\p S} F \cdot ds.$$ 
\newpage

\section*{Appendix B}\label{appendixB}
 \subsection{Relation to previous versions}

Chainlets have been developing over a period of years.  In earlier versions, they relied on coordinate expressions which made proofs long and relatively cumbersome. They initially took the viewpoint of Whitney, starting with completions of polyhedral chains with respect to a norm.  The Ravello lecture notes \cite{ravello}  started with polyhedral chains and developed the first discrete chainlet theory with the definition of a {\itshape   $k$-element} as a geometric representation of a $k$-vector, based at a point, now called a {\itshape monopolar $k$-vector}.  A {\itshape   monopolar $k$-chain}  is a finite sum of monopolar $k$-vectors supported in finitely many points, and these form a dense subset of chainlets, as seen in \cite{ravello}. The main object of the Berkeley lecture notes \cite{gmt} was to produce a coordinate free theory of chainlet calculus without any assumption of the integral theorems of classical calculus.

The initial goal of this paper was to develop the theory without using polyhedral chains, except as examples.  The Koszul complex aided enormously in this quest.  Setting aside polyhedral chains soon led to large simplifications of the theory.  Now there is essentially one classical limit to establish for the full calculus, beyond the Cauchy sequences needed for examples, namely that of Lemma \ref{mainlimit} proving the existence of the prederivative operator.

\section*{Conclusion}
The prederivative operator $\nabla_u$ is an operator at the very foundations of mathematics. For $\nabla_u$ leads to boundary, which, in turn leads to exterior derivative through duality.  It links boundary $\p$  to exterior product  $e_u(\a) = (\a \wedge u)$ in  a ``magic formula'' $$\nabla_X J  = \p e_X J  + e_X \p J.$$      The operator $\nabla_u$ leads to the existence of norms on chainlets which, in turn, leads to uniform convergence theorems, not available for distributions and currents.   The prederivative, combined with the preintegral, leads to a general divergence theorem with three integrals, rather than two, that models well conservation of matter and energy in curved space.  The full continuum with its polypoles of all dimensions and orders reveals mysteries of mathematics and physics. Both simple and multilayered complex systems naturally arise from this which are self-organizing. The polypoles are natural candidates for the basic components of a ``genetic code'' of the universe itself.

\end{document}